\documentclass[journal,twoside,web]{ieeecolor}
\usepackage{tuffc}
\usepackage{cite}
\usepackage{amsmath,amssymb,amsfonts}
\usepackage{algorithmic}
\usepackage{graphicx}
\usepackage{textcomp}
\usepackage{wrapfig,colortbl}

\usepackage{upgreek}
\usepackage{hyperref}
\usepackage{multirow}
\usepackage{tabularx}
\definecolor{abstractbg}{rgb}{1,0.969,0.914}
\def\BibTeX{{\rm B\kern-.05em{\sc i\kern-.025em b}\kern-.08em
    T\kern-.1667em\lower.7ex\hbox{E}\kern-.125emX}}
    
\begin{document}
\onecolumn
This work will be submitted to the IEEE for possible publication. Copyright may be transferred without notice, after which this version may no longer be accessible.

\twocolumn
\newpage
\title{Synthetic Aperture for High Spatial Resolution Acoustoelectric Imaging}
\author{Wei Yi Oon, \IEEEmembership{Student Member, IEEE}, Yuchen Tang, \IEEEmembership{Member, IEEE}, Baiqian Qi, \IEEEmembership{Student Member, IEEE}, and Wei-Ning Lee, \IEEEmembership{Senior Member, IEEE}
\thanks{Manuscript submitted for review XX XXX 2025. Wei Yi Oon is supported by the Hong Kong PhD Fellowship Scheme and The University of Hong Kong Presidential Scholarship. This work was supported by the Seed Fund (basic research) of the University of Hong Kong (project number: 104005865).}
\thanks{Wei Yi Oon, Yuchen Tang and Baiqian Qi are with the Department of Electrical and Electronic
Engineering, The University of Hong Kong, Hong Kong.}
\thanks{Wei-Ning Lee is with the Department of Electrical and Electronic
Engineering and School of Biomedical Engineering, The University of
Hong Kong, Hong Kong (e-mail: wnlee@hku.hk).}}

\IEEEtitleabstractindextext{%
\fcolorbox{abstractbg}{abstractbg}{%
\begin{minipage}{\textwidth}\rightskip2em\leftskip\rightskip\bigskip
\begin{wrapfigure}[15]{r}{3in}%
\hspace{-3pc}\includegraphics[width=2.9in]{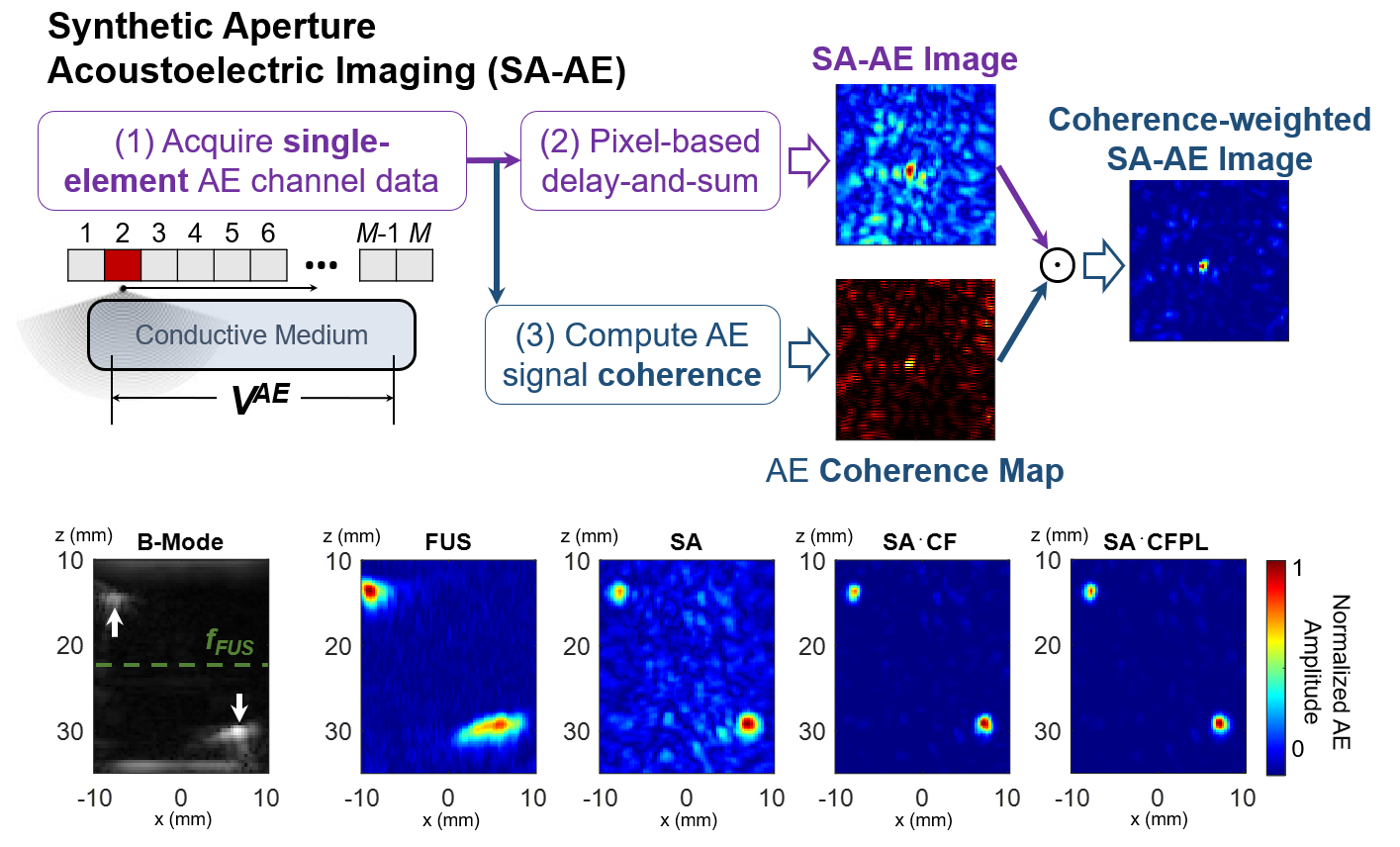}
\end{wrapfigure}%
\begin{abstract}
Acoustoelectric (AE) imaging provides electro-anatomical contrast by mapping the distribution of electric fields in biological tissues, by delivering ultrasound waves which spatially modulate the medium resistivity via the AE effect. The conventional method in AE imaging is to transmit focused ultrasound (FUS) beams; however, the depth-of-field (DOF) of FUS-AE is limited to the size of the focal spot, which does not span across the centimeter-scale of organs. Instead of fixing the focal depth on transmission, we propose to dynamically synthesize the AE modulation regions via a Synthetic Aperture approach (SA-AE). SA-AE involves a straightforward pixel-based delay-and-sum reconstruction of AE images from unfocused AE signals. In saline and \textit{ex vivo} lobster nerve experiments, FUS-AE was shown to perform well only at the focal depth, with poor spatial resolution for out-of-focus electric sources. Meanwhile, SA-AE generally improved spatial resolution throughout the DOF, but introduced strong background noise. The flexibility of uncoupled, single-element induced AE signals in SA-AE was further leveraged to quantify their spatial coherence across the transmit aperture, obtaining maps of the coherence factor (CF) and pulse-length coherence factor (CFPL). Weighting SA-AE images with their derived CF and CFPL maps resulted in further improvement in image resolution and contrast, and notably, boosted the image SNR beyond that of FUS-AE. CFPL exhibited stronger noise suppression over CF. Using unfocused wave transmissions, the proposed coherence-weighted SA-AE strategy offers a high resolution yet noise-robust solution towards the practical imaging of fast biological currents.
\end{abstract}

\bigskip
\end{minipage}}}

\maketitle

\begin{table*}[!t]
\arrayrulecolor{subsectioncolor}
\setlength{\arrayrulewidth}{1pt}
{\sffamily\bfseries\begin{tabular}{lp{6.75in}}\hline
\rowcolor{abstractbg}\multicolumn{2}{l}{\color{subsectioncolor}{\itshape
Highlights}{\Huge\strut}}\\
\rowcolor{abstractbg}$\bullet$ & The proposed SA-AE approach dynamically synthesized pixel-based AE modulation regions, thereby extending the depth-of-field of AE imaging without requiring a prescribed focal depth.\\
\rowcolor{abstractbg}$\bullet${\large\strut} & Compared to FUS-AE, SA-AE imaging improved spatial resolution consistently throughout the depth-of-field, and coherence-based weighting boosted the SNR performance of SA-AE.\\
\rowcolor{abstractbg}$\bullet${\large\strut} & The demonstration of SA-AE shows the value of considering the design of beam weighting functions as a general approach in AE imaging, to provide application-based optimization of AE image quality.\\[2em]\hline
\end{tabular}}
\setlength{\arrayrulewidth}{0.4pt}
\arrayrulecolor{black}
\end{table*}

\section{Introduction}
\label{sec:introduction}

\IEEEPARstart{A}{coustolectric} (AE) imaging refers to the mapping of electric fields with the use of ultrasound (US) waves as the excitation source. The technique is based on the AE effect, which describes the modulation of medium conductivity due to the compressional action of acoustic waves \cite{fox1946effect}. Enabled by remote insonification and deep penetration of US waves, AE imaging holds potential for non-invasive mapping of electrical contrast in biological media, such as tissue resistivity \cite{zhang2004acousto} and biological currents in excitable tissues like the heart and brain \cite{alvarez2020vivo, zhou2024vivo}. Conventional surface electro-physiological mapping modalities require the use of a large number of electrodes to ensure sensitivity, or otherwise rely on the inversion of ill-posed problems to solve for an electric field distribution from a limited number of electrode pairs \cite{ruiz2025body,farina2023evolution}. Meanwhile, owing to the sifting ability of the US beam, AE imaging can achieve direct full-field electric contrast mapping from only a few electrode pairs \cite{olafsson2008ultrasound}.

The magnitude of the lead-measured AE signal is determined by the strength of the projected electric field and the US beam pattern applied to the system \cite{olafsson2008ultrasound}. The design of AE imaging strategies involves the tuning of US transmission schemes to establish the beam patterns needed for targeted probing of the electric fields within the medium. Commonly, such beams are delivered by applying focused ultrasound (FUS) waves \cite{olafsson2008ultrasound,alvarez2020vivo,allard2024neuronavigation}. FUS waves concentrate acoustic energy at a prescribed focal spot, thereby sifting the electric properties at that point. While FUS is effective in the localization of AE sources, they have a limited depth-of-field (DOF) due to its finite focal size. At diagnostic US MHz frequencies, the focal spot is typically a only few mm in size. Therefore, the range of focused ultrasound AE (FUS-AE) images are constrained to around the insonified focal depth, and are unable to cover organs spanning a larger DOF.

Approaches which depart from the use of FUS waves for AE imaging have been demonstrated. Plane wave emissions have been employed to achieve ultrafast AE imaging (UAI), showing improved contrast and spatial resolution compared to FUS-AE \cite{berthon2017integrated}. However, the UAI image formation procedure is elaborate, requiring prior simulation of the full-field waveforms to establish an inversion matrix for reconstruction. Meanwhile, Hadamard-encoded transmissions have been shown to maximize the field-of-view while preserving spatial resolution \cite{preston2022hadamard}. Although these studies showcase the value of unfocused transmission schemes for AE imaging, they lack a direct discussion on how the changes on the transmitted beam pattern effectively influences the formation of AE images.

To provide a more comprehensive view on the role of beamforming in AE imaging, we revisit the AE image reconstruction problem using the general scheme of transmitting unfocused US waves. Synthetic aperture (SA) ultrasound uses single-element transmissions of spherical waves, in which the independently induced returning echos are dynamically combined to reconstruct B-mode images of theoretically ideal spatial resolution \cite{jensen2006synthetic}. Inspired by this approach, we show that AE images can be similarly formed following a straightforward DAS approach, to achieve synthetic beam focusing across the entire DOF. We recognize that the principle of SA-based AE (SA-AE) imaging has been indirectly realized as a substituent part of AE reconstruction from Hadamard-encoded transmissions \cite{preston2022hadamard}, but a noticeable background noise issue remains, which may explain its limited use in practice.

In this work, our aim is to show how several beamforming approaches can improve AE imaging performance in various aspects, with the central theme of ``beam weighting function determines AE image structure". Starting from SA-AE, we show that a dynamically synthesized focus helps to establish sifting beams throughout the DOF, to preserve a relatively uniform spatial resolution. At the same time, we present the effects of performing FUS-AE imaging on electric sources with an unknown depth position to show how an imperfect beam pattern distorts the AE image. By considering the noise characteristics in AE signals, we show the dependence of AE image signal-to-noise ratio (SNR) on the number of transmit elements and temporal averages used. In doing so, we explain why SA-AE has an inherent disadvantage in SNR to FUS-AE while maintaining a fixed frame rate. As SA-AE reconstructed images are prone to background noise, we further supplemented this technique with coherence-based beamforming to suppress possible focusing artifacts and incoherent background noise. Overall, the purpose of this study is to elucidate the link between the effective beam patterns used and the resulting AE imaging quality, while using SA-AE as an example to show how multiple beam shaping strategies can improve its feasibility for practical applications.

\section{Theory}
\label{sec:theory}

\subsection{Role of Beam Pattern in AE Signal Generation}

Considering a conductor domain $D$, the general temporal AE signal in 2D \cite{olafsson2008ultrasound} can be expressed as 
\begin{IEEEeqnarray}{c}
V^{AE}(t) = -K_I\int_D s(\textbf{x})\Delta P(\textbf{x},t)\,dA,
\label{eq:vae_gen}
\end{IEEEeqnarray}
where \textbf{x} is the spatial coordinate vector for point ($x,z$) within $D$, $t$ is the US fast-time (s), $\Delta P$ is the pressure field of the applied US wave (Pa), and $K_I$ is the AE interaction constant \cite{jossinet1999impedance} quantifying the fractional change in resistivity per unit pressure change (Pa$^{-1}$). $s(\textbf{x})$ represents the projected electric field measured by a particular lead, which consists of the following sub-components:
\begin{IEEEeqnarray}{c}
s(\textbf{x})=\textbf{J$^L$}(\textbf{x})\cdot(\rho_0(\textbf{x})\textbf{J$^I$}(\textbf{x})),
\label{eq:s}
\end{IEEEeqnarray}
where $\rho_0$ is the local medium resistivity ($\Omega\cdot$m), \textbf{J$^L$} and \textbf{J$^I$} are the measurement lead field (m$^{-2}$) and 2D current density field (A$\cdot$m$^{-1}$), respectively. 
The mapping target in AE imaging is this $s$ field, as it encapsulates both the underlying impedance distribution \cite{zhang2004acousto} and current density fields \cite{olafsson2008ultrasound}. In general, the pressure field, $\Delta P$, can be decomposed as
\begin{IEEEeqnarray}{rl}
\Delta P(\textbf{x},t) = P_0b(\textbf{x})a(t-|\textbf{r}|/c)
\label{eq:p_xyz}
\end{IEEEeqnarray}
where $P_0$ describes the pressure amplitude, $b(\textbf{x})$ the US transmit beam pattern, $a(t)$ the finite-duration pulse waveform centred at $t=0$, and $c$ the speed of sound (m$\cdot$s$^{-1}$). \textbf{r} refers to the displacement vector between each spatial point and the US source ($\textbf{r} = \textbf{x}-\textbf{x$_s$}$), hence, $|\textbf{r}|/c$ is the time-of-arrival of the pulse at each point \textbf{x}. Combining \eqref{eq:vae_gen} and \eqref{eq:p_xyz} yields
\begin{IEEEeqnarray}{rl}
V^{AE}(t) = -K_IP_0\int_D & s(\textbf{x})b(\textbf{x})a(t-|\textbf{r}|/c)\,dA,
\label{eq:vae_beam}
\end{IEEEeqnarray}
which highlights the role of $a(t-|\textbf{r}|/c)$ and $b(\textbf{x})$ as temporal and spatial weightings on the targeted $s(\textbf{x})$ field.

AE imaging aims to reconstruct the $s$ field by mapping the detection times of AE signals to their corresponding spatial positions in the medium. Conventional B-mode US is pulse-echo imaging. Receive beamforming can be done in a pixel-wise manner, which calculates round-trip time of the US waves. For a pixel of interest, the transmission path counts the travel time from a wave source to the target pixel; the reception path considers the return time of the echo from the target pixel to each active element in a receive aperture, thereby allowing spatial localization of acoustic signals. However, in AE imaging, received signals are electrical signals that practically propagate instantaneously across the cm-scale imaging domains. Even if a spatially distributed electrode array is employed, the AE signals induced at a given source will be detected at the same time across the electrode array. Hence, selection of signal samples by applying spatially-corresponding receive delays is impossible. For this reason, AE imaging of the $s$ field leverages the US propagation trajectory for spatial mapping utilizing transmit delays only, and is essentially a one-way imaging mode.

For a given point $\textbf{x$_1$}=(x_1,z_1)$, the apparent $s$ value, $\hat{s}$, can be extracted by first applying the source-to-point transmit time delay of $t_1=|\textbf{r$_1$}|/c$ into \eqref{eq:vae_beam}:
\begin{IEEEeqnarray}{rl}
\hat{s}(\textbf{x$_1$}) = -K_IP_0\int_D s(\textbf{x})b(\textbf{x})a(\frac{|\textbf{r$_1$}|-|\textbf{r}|}{c})\,dA.
\label{eq:shat_D}
\end{IEEEeqnarray}
Note that without receive beamforming, \eqref{eq:shat_D} provides only partial spatial selectivity, as it contains contributions of the $s$ field for all points sharing the same transmit delay $t_1$. For the case of a spherical wave, this band of iso-temporal points, $I_{t_1}$, lies on the spherical surface of the transmitted wavefront. To describe the extent of temporal sifting obtained through the application of transmit delays, we model $a(t)$ as
\begin{IEEEeqnarray}{c}
a(t)=\delta(t),
\label{eq:a_t}
\end{IEEEeqnarray}
where $\delta$ is the Dirac delta function. Equation \eqref{eq:a_t} assumes an US impulse, which sifts the local $s(\textbf{x$_1$})$ field only when the pulse arrives at \textbf{x$_1$}. Substituting \eqref{eq:a_t} into \eqref{eq:shat_D}, we obtain
\begin{IEEEeqnarray}{rl}
\hat{s}(\textbf{x$_1$}) = -K_IP_0\int_{I_{t_1}} & s(\textbf{x})b(\textbf{x})\,dA,
\label{eq:shat_I}
\end{IEEEeqnarray}
expressing the reduction of the integration domain from $D$ to the band of iso-temporal points $I_{t_1}$. Hence, simply applying the transmit delay $t_1$ only narrows down signal sources to the iso-temporal points, but does not pinpoint the signal to \textbf{x$_1$} yet.

In order for $\hat{s}(\textbf{x$_1$})$ to sample $s(\textbf{x$_1$})$ accurately, the weighting provided by the beam at the desired point, $b(\textbf{x$_1$})$, must be large enough to outweigh the other points in $I_{t_1}$. In other words, the main lobe of the beam pattern should be as narrow as possible. Suppose that we have some $b(\textbf{x$_1$})$ which fulfills this criterion, then \eqref{eq:shat_I} is approximate to
\begin{IEEEeqnarray}{c}
\hat{s}(\textbf{x$_1$}) \approx -K_IP_0b(\textbf{x$_1$})s(\textbf{x$_1$}).
\label{eq:shat_x1_approx}
\end{IEEEeqnarray}
Such a beam allows $\hat{s}(\textbf{x$_1$})$ to sift out $s(\textbf{x$_1$})$ from the $s$ field, while being amplified by the local beam strength $b(\textbf{x$_1$})$.

Equations \eqref{eq:shat_I} and \eqref{eq:shat_x1_approx} show that $b$ plays a crucial role as a weighting function in defining the AE modulation region, and therefore in the mapping of $s$. First, a pixel-targeted $b(\textbf{x})$ provides a narrow point spread function, allowing local modulation of the signal source in AE imaging. Second, $b(\textbf{x$_1$})$ controls the degree of amplification on $s(\textbf{x$_1$})$, and hence the AE signal-to-noise ratio (SNR) measured. At the same time, the apparent magnitude of the reconstructed signal $\hat{s}(\textbf{x$_1$})$ is determined by the scaling provided by $b(\textbf{x$_1$})$, so the uniformity of AE image contrast is also dictated by the uniformity of applied local beam weighting function.

\subsection{Conventional Focused Ultrasound AE Imaging}

The common approach for achieving spatial selectivity in AE imaging is by delivering FUS waves, which sift out the $s$ field at the prescribed beam focus \cite{olafsson2008ultrasound,alvarez2020vivo,zhou2024vivo}. For the case of a focused beam positioned at $x=x_1$, \eqref{eq:vae_beam} is rewritten as
\begin{IEEEeqnarray}{rl}
V^{AE}_{F}(x_1,t) = -K_IP_0\int_D & s(\textbf{x})b_f(x-x_1,z)a(t-\frac{z}{c})\,dA,
\label{eq:vae_FUS}
\end{IEEEeqnarray}
where $b_f(x,z)$ is the focused beam pattern centered at $x_1=0$ with its focal depth at $z=f$. Like B-mode focused rayline imaging, FUS-AE assumes that the signal source resides within the beam axis ($x=x_1$), so time $t$ only reflects depth information in the $z$-direction. In this case, $\hat{s}$ at $\textbf{x$_1$}=(x_1,z_1)$ is mapped by applying the transmit time delay of $t_1=z_1/c$:
\begin{IEEEeqnarray}{l}
\hat{s}_{F}(\textbf{x$_1$}) = -K_IP_0\iint s(\textbf{x})b_f(x-x_1,z)a(\frac{z_1-z}{c})\,dx\,dz.
\label{eq:shat_FUS}
\end{IEEEeqnarray}
We further consider a narrow beam width, i.e.,
\begin{IEEEeqnarray}{c}
b_f(x,z) = \delta(x)b_z(z),
\label{eq:b_xz}
\end{IEEEeqnarray}
where $\delta$ is the Dirac delta function and $b_z(z)$ is the axial beam profile. Note that this assumption is valid only within the focal zone, at around $z=f$. Accounting for the finite focal extent spanning $f_s\le z\le f_e$ and applying the sifting properties of \eqref{eq:a_t} and \eqref{eq:b_xz}, \eqref{eq:shat_FUS} is expanded as
\begin{IEEEeqnarray}{l}
\hat{s}_{F}(\textbf{x$_1$})
\nonumber\\
= \left\{\begin{array}{ll}
-K_IP_0b_z(z_1)s(\textbf{x$_1$}), & \text{for }f_s\le z_1\le f_e;\\
-K_IP_0\int s(x,z_1)b_f(x-x_1,z_1)\,dx, &
\text{otherwise.} \\
\end{array}\right.
\label{eq:shat_FUS_exp}
\end{IEEEeqnarray}
From \eqref{eq:shat_FUS_exp}, 
it can be seen that the FUS beam weighting function provides lateral sifting effectively only within the focal zone. Beyond that, the measured AE signal is less spatially specific and instead captures the $s$ field from a broader band of off-axis points. The validity of FUS-AE images is therefore limited to $f_s\le z\le f_e$, due to the fixed focused beam patterns in $b_f(x,z)$ applied during the imaging process.

For this reason, FUS-AE imaging requires prior knowledge of the region of interest within the imaging field-of-view to enable the selection of a suitable focal depth $f$ \cite{alvarez2020vivo,allard2024neuronavigation}. While this method is effective for acquiring AE images of electrical sources around the focal depth, its depth-of-field (DOF) is constrained to the fixed focal zone used. As such, conductive domains which span beyond the size of the focal extent cannot be imaged completely using standard FUS-AE imaging.

An ideal extension of the DOF should allow effective sifting of the $s$ field continuously throughout the imaging domain, rather than at discrete focal spots. To achieve this goal, the weighting functions ($b(\textbf{x})$ in \eqref{eq:shat_I}) applied in the reconstruction of each image point should exhibit high spatial specificity; yet, their overall sifting effects should be spatially uniform to allow accurate depiction of the true $s$ field distribution. In other words, in contrast to FUS-AE imaging, which involves one beam per axial line, each AE image pixel should be reconstructed from its own dynamically focused beam.

\subsection{Synthetic Aperture Beamforming for AE Imaging}

Like in B-Mode SA \cite{jensen2006synthetic}, transmit focusing can be synthesized for AE imaging by sequential single-element transmissions, followed by pixel-based Delay-And-Sum (DAS). We term this technique as SA-AE imaging. To synthesize a transmit beam pattern focused at \textbf{x$_1$}, the delayed signal, $\hat{s_i}(\textbf{x$_1$})$ in \eqref{eq:shat_I}, for each single-element transmit $i$ is coherently summed across the full transmit aperture consisting of $M$ elements:
\begin{IEEEeqnarray}{rl}
\hat{s}_{SA}(\textbf{x$_1$})\, = \sum_{i=1}^{M} \hat{s_i}(\textbf{x$_1$}) = -K_IP_0\sum_{i=1}^{M}\left(\int_{I_{t_1}^i} s(\textbf{x})b_i(\textbf{x})\,dA\right).
\label{eq:shat_SA}
\end{IEEEeqnarray}
Note that each transmit $i$ is associated with a unique arc of iso-temporal points, $I_{t_1}^i$, because each transmission sources from a different lateral position. $b_i$ denotes the shifted beam patterns from each single-element transmit. Since the transmit delays are calculated to target \textbf{x$_1$}, the $M$ sets of $I_{t_1}^i$ intersect at \textbf{x$_1$} only. Consequently, this coherent summation results in the synthesis of an effective beam weighting function, $b_{\textbf{x$_1$}}$, which is preferentially focused at \textbf{x$_1$}, among points across all $I_{t_1}^i$. Rewriting \eqref{eq:shat_SA} to illustrate this synthesized beam, we obtain
\begin{IEEEeqnarray}{rl}
\hat{s}_{SA}(\textbf{x$_1$})\, & = -K_IP_0\int s(\textbf{x})b_{\textbf{x$_1$}}(\textbf{x})\,dA
\label{eq:shat_SA_bSA}
\\
& \approx -K_IP_0b_{\textbf{x$_1$}}(\textbf{x$_1$})s(\textbf{x$_1$}).
\label{eq:shat_SA_approx}
\end{IEEEeqnarray}
In fact, \eqref{eq:shat_SA_bSA} represents AE imaging with a beam focused at \textbf{x$_1$}, analogous to FUS-AE. The main difference, emphasized by the subscript \textbf{x$_1$} in $b_{\textbf{x$_1$}}$, is that each spatial point relies on a distinct beam weighting function synthesized for its own reconstruction. Since focusing is performed dynamically for all pixels, the full SA-AE image effectively sees a point spread function that is more uniform than FUS-AE across the entire DOF. The approximation in \eqref{eq:shat_SA_approx} holds when the focusing power at $b_{\textbf{x$_1$}}(\textbf{x$_1$})$ is large enough to sift out $s(\textbf{x$_1$})$ from the integral in \eqref{eq:shat_SA_bSA}. In such cases, \eqref{eq:shat_SA_approx} shows that dynamic synthesis of $b_{\textbf{x$_1$}}$ produces a point-specific sifter of the $s$ field at \textbf{x$_1$}, aligning with the ideal case presented in \eqref{eq:shat_x1_approx}. 

Additionally, the independently acquired single-element signal channels in SA-AE allow for flexible selection of depth-dependent transmit sub-apertures during reconstruction. The transmit F-number, $F_n$, can be fixed to enforce uniform focusing power throughout the DOF. To implement a fixed $F_n$ for SA-AE, the coherent summation in \eqref{eq:shat_SA} is modified to include only the $M_{sa}$ elements which cover the sub-aperture width required at each depth $z$. $M_{sa}$ is defined as
\begin{IEEEeqnarray}{c}
M_{sa}(z) = \text{round}(z/(F_np)),
\label{eq:M_sub}
\end{IEEEeqnarray}
where $F_n$ is the chosen F-number, $p$ is the array element pitch and ``round" is the nearest-integer rounding function. Unlike SA-AE, FUS-AE lacks this transmit sub-aperture flexibility since the number of elements is fixed upon transmission and cannot be decoupled into independent AE signal channels.

\subsection{AE SNR Dependency on Beamforming and Averaging}
\label{subsection:Theory_SNR}

In the analog amplification of raw AE signals by their front-end instrumentation, the signals are inevitably contaminated with electronic thermal noise. Due to the $\upmu$V-range amplitude of AE signals, thermal noise is a non-negligible noise component in raw AE data, similar to the dominance of thermal noise in US radio-frequency data in deep tissue regions \cite{goudarzi2023deep}, where returning echoes are markedly weak. Thermal noise can be modelled by additive Gaussian noise \cite{haynes2019homodyned}, which exhibits a variance that scales with $1/k$ when $k$ independent Gaussian noise samples are averaged \cite{papoulis1965random}. Therefore, to ensure sufficient SNR, AE signal acquisition requires coherent signal averaging to reduce noise in each voltage measurement. Using $k$-transmit averaging, a raw voltage temporal profile ($V^{AE}(t)$ in \eqref{eq:vae_beam}) is measured by averaging the electrical signal coherently across the pulse propagation duration, over $k$ repetitions of the same US transmission. This results in a reduction of noise power by $1/k$, but lowers the frame rate by $1/k$ due to the repeated transmission propagation times.

Consider one channel of raw SA-AE signal induced by a single transmit element and averaged over $k$ repetitions. Its SNR, $\text{SNR}_{1,k}$ can be represented as
\begin{IEEEeqnarray}{rl}
\text{SNR}_{1,k} = S_\phi/(N/k)=k(S_\phi/N), 
\label{eq:SNR_SE}
\end{IEEEeqnarray}
where $S_\phi$ is the AE signal power induced by one transmit element for a given electric source $\phi$, and $N$ is the background thermal noise power without any temporal averaging. In the SA-AE DAS reconstruction, the summing of delayed signals from $M_{sa}$ channels amplifies the signal and noise powers by a factor $M_{sa}{}^2$ and $M_{sa}$, respectively, resulting in an SNR of
\begin{IEEEeqnarray}{rl}
\text{SNR}_{SA,k} = (M_{sa}{}^2S_\phi)/M_{sa}N/k) = kM_{sa}(S_\phi/N).
\label{eq:SNR_SA}
\end{IEEEeqnarray}
To measure a fair SNR comparison for FUS-AE, we quantify its SNR based on the same source, $\phi$, which should be located at the focal point of the FUS transmission. Assuming linear superposition of acoustic pressure, the FUS-induced AE signal power would be amplified by $M_f{}^2$, where $M_f$ is the number of transmit elements used for FUS transmission. Therefore, the SNR for FUS-AE with $k$ averages, $\text{SNR}_{F,k}$, is:
\begin{IEEEeqnarray}{rl}
\text{SNR}_{F,k} = (M_f{}^2S_\phi)/(N/k) = kM_f{}^2(S_\phi/N).
\label{eq:SNR_FUS}
\end{IEEEeqnarray}
Dividing \eqref{eq:SNR_SA} by \eqref{eq:SNR_FUS} and rearranging, we get:
\begin{IEEEeqnarray}{rl}
\text{SNR}_{SA,k} = (M_{sa}/M_f{}^2)\cdot\text{SNR}_{F,k},
\label{eq:SNR_SAFUS}
\end{IEEEeqnarray}
showing that the SNR of SA-AE is suppressed by $M_{sa}/M_f{}^2$ compared to FUS-AE. Note that the relative SNR performance in \eqref{eq:SNR_SAFUS} applies when comparing the SNR at an electric source, $\phi$, located at the FUS focus. For sources outside the FUS focus, the SNR gap between the two techniques will diminish, since imperfect focusing would reduce $\text{SNR}_{F,k}$; while $\text{SNR}_{SA,k}$ is relatively unaffected due to its dynamic focusing.

\subsection{Further Improvement for SA-AE Imaging}

Despite their spatial resolution enhancement, SA schemes suffer from reduced SNR due to the low acoustic energies from single-element transmissions \cite{o2002efficient}. The additional SNR reduction inherent to the SA approach (presented in \eqref{eq:SNR_SAFUS}) restricts practical applications of SA-AE imaging. To compensate for this reduction in SNR, one could increase the the number of averages for SA-AE to $k_2$ as follow:
\begin{IEEEeqnarray}{rl}
k_2 = (M_f{}^2/M_{sa})k.
\label{eq:k2}
\end{IEEEeqnarray}
Then, substituting (\ref{eq:k2}) into (\ref{eq:SNR_SA}), we obtain:
\begin{IEEEeqnarray}{rl}
\text{SNR}_{SA,k_2} = ((M_f{}^2/M_{sa})k)M_{sa}(S_\phi/N) = \text{SNR}_{F,k}.
\label{eq:SNR_SAk2}
\end{IEEEeqnarray}
Equation (\ref{eq:SNR_SAk2}) shows that for SA-AE to reach the same SNR level as FUS-AE, the number of averages must be scaled by $M_f{}^2/M_{sa}$. This factor is on the order of the number of transducer array elements $M$, which typically ranges between $10^1-10^3$. A brute-force increase in averaging for boosting SA-AE SNR would severely decrease the frame rate of SA-AE, hindering its applicability towards imaging transient bio-currents. To circumvent the reliance on additional averaging for satisfactory SA-AE image quality, we additionally explore non-linear coherence factor beamforming to enhance SA-AE spatial resolution, contrast and SNR.

The fact that electronic noise can be reduced by averaging indicates that it is an incoherent component contained within the set of AE voltage signals. Therefore, we seek to delineate pixels that contain coherent signal content reflecting the $s$ field, from pixels corrupted by incoherent electronic noise. In contrast to the FUS-AE data, the SA-AE channel data allow quantification of spatial coherence since a given electrical source in the $s$ field is insonified by isolated single-element transmissions. We adopt the conventional Coherence Factor (CF) \cite{hollman1999coherence} to quantify this AE signal coherence. For all pixels, we compute the pixel-based CF from the transmit-delayed SA-AE aperture data ($\hat{s_i}(\textbf{x})$ in \eqref{eq:shat_SA}) as follow:
\begin{IEEEeqnarray}{c}
\text{CF}(\textbf{x})=\frac{\left|\sum_{i\in S}\hat{s_i}(\textbf{x})\right|^2}{M_{sa}\sum_{i\in S}|\hat{s_i}(\textbf{x})|^2}
\label{eq:CF}
\end{IEEEeqnarray}
As in \eqref{eq:M_sub}, $M_{sa}$ depends on the width of the dynamic sub-aperture used for SA-AE reconstruction. $S$ is the set of element indices covering the sub-aperture of $M_{sa}$ elements.
We further considered that the modulation effect from each pulse has a finite temporal extent, due to their finite pulse lengths. An extended ``pulse length" CF (CFPL) was computed as follow: 
\begin{IEEEeqnarray}{c}
\text{CFPL}(\textbf{x})=\frac{1}{P}\cdot\sum^{P}_{j=1}\left(
\frac{\left|\sum_{i\in S}\tilde{s_i}(\textbf{x},t_j)\right|^2}{M_{sa}\sum_{i\in S}|\tilde{s_i}(\textbf{x},t_j)|^2}
\right)
\label{eq:CFPL}
\end{IEEEeqnarray}
where $P$ is the total temporal samples equivalent to the pulse length. $\tilde{s_i}(\textbf{x},t_j)$ is similar to 
$\hat{s_i}(\textbf{x})$ in \eqref{eq:shat_SA} and \eqref{eq:CF}, except that instead of acquiring the single DAS sample at the time-of-arrival, the $P$ samples spanning the pulse length were acquired, to obtain a temporal DAS profile for each pixel. The coherence factor at each time instant is computed identically as in \eqref{eq:CF}, followed by a temporal mean across the $P$ samples. The computed CF and CFPL maps were then pixel-wise multiplied with the beamformed SA-AE signals.

\section{Methods}
\label{sec:methods}

\subsection{Instrumentation for AE Signal Measurement}

\begin{figure}[!t]
\centerline{\includegraphics[width=0.85\columnwidth]{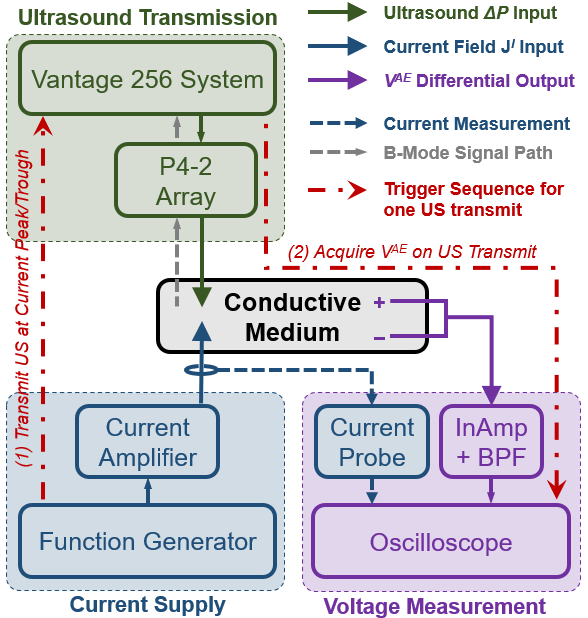}}
\caption{Instrumentation block diagram with AE signal acquisition trigger sequence. The setup consists of three main parts: (1) an US transmission module to provide the US pressure field $\Delta P$, (2) a current supply module to establish the current field to be imaged, (3) a voltage measurement module to record the induced AE signals $V^{AE}$. The red dot-dash arrows indicate the trigger sequence between the modules during signal acquisition under one US transmission. With $k$-transmit averaging, this loop was repeated $k$ times to measure one AE data channel. InAmp: Instrumentation amplifier, BPF: Bandpass filter.}
\label{fig_instrumentation}
\end{figure}

The instrumentation used to acquire the AE voltage signals is shown in Fig. \ref{fig_instrumentation}.
A P4-2 array (Philips Healthcare, Andover, MA, USA) was used with a Vantage 256 system (Verasonics, Kirkland, WA, USA) to transmit 1-cycle US pulses ($f_c=2$ MHz) at an US driving voltage of 50V for AE signal induction. Sinusoidal currents at 1 kHz from a signal generator (33320A; Agilent, Santa Clara, CA, USA) were fed through an amplifier (OPA547; Texas Instruments, Dallas, TX, USA) with a gain of 12 dB, before being injected into the conductive phantoms using Pt wire electrodes (0.5 mm diameter). The current flowing through the supply electrodes was measured using a current probe (N2783B; Keysight, Santa Rosa, CA, USA). On receive, differential voltage measurements were made using a Pt electrode pair, which were fed through a high-pass instrumentation amplifier circuit (AD8428; Analog Devices, Wilmington, MA, USA), followed by a 2-7 MHz bandpass filter (ZABP-4R5-S+; Mini-Circuits, Brooklyn, NY, USA). The output signal was measured by an oscilloscope (DSOX3024T; Keysight, Santa Rosa, CA, USA). The receiving circuits provided an overall gain of 38 dB.

In this work, we primarily compared between the FUS-AE and SA-AE imaging strategies. In FUS-AE, all 64 elements of the P4-2 array were active for transmission, and the focused beams were swept across the array to produce 64 AE ray lines. The FUS transmit delay profile was calculated based on a speed-of-sound (SoS) of 1480 m$\cdot$s$^{-1}$. B-mode images of the conductive bodies were also acquired, using the same set of FUS ray lines and a full receive aperture. Meanwhile, SA-AE uses one active transmit element per acquisition, for a total of 64 single-element AE signal channels.

\subsection{AE Data Acquisition Sequence and Processing}

The red dot-dash arrows in Fig. \ref{fig_instrumentation} show the trigger events for one US pulse transmission during signal acquisition. In each acquisition cycle, the signal generator sent the master trigger, which instructed: (1) the Vantage 256 system to transmit the US pulse at the positive peak of the current cycle; and (2) the oscilloscope to begin AE voltage acquisition for the pulse transmitted. The AE signal was sampled at 385 MHz for the duration of one-way propagation across the field-of-view.

\begin{figure}[!t]
\centerline{\includegraphics[width=0.85\columnwidth]{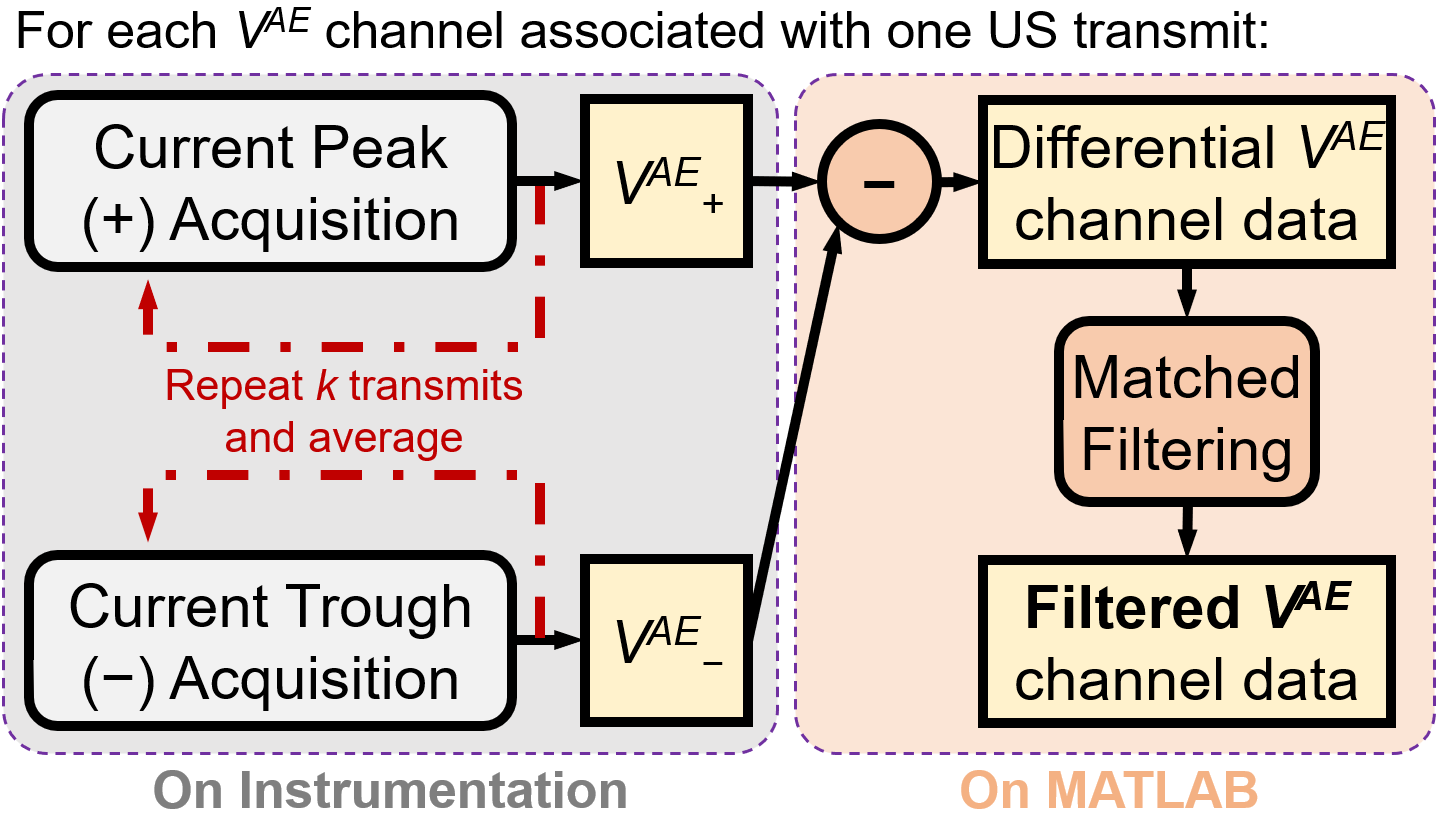}}
\caption{Data acquisition flowchart for AE channel data. For a given transmit pattern, the AE signal was measured at the current peak and trough (with $k$ repeated transmissions each). The ($+$) and $(-)$ temporal data were then subtracted and matched filtered to obtain a single channel of filtered pre-beamformed $V^{AE}$ data.}
\label{fig_acquisition}
\end{figure}

For each data channel associated with one transmit pattern (Fig. \ref{fig_acquisition}), the induced temporal AE signal was averaged over $k$ repeated transmissions, and then recorded to obtain one positive raw data channel. After all 64 data channels had been acquired (corresponding to 64 ray lines in FUS-AE and 64 single-element transmits in SA-AE), the procedure above was repeated, sampling the AE signal at the peak negative current in place of the positive peak. The subsequent steps were performed on MATLAB. For common-mode noise reduction, the AE channel data acquired at the positive peak was subtracted from that at the negative peak to obtain the raw differential AE data channels. Matched filtering was performed by convolving the raw differential data with the emitted acoustic pulse waveform (simulated by the Vantage 256 software). The amplitude of the pulse template was scaled such that the convolution procedure would not alter the voltage amplitude of the AE signals. Subsequent computation of AE images used these filtered differential AE signals as the pre-beamformed channel data.

\subsection{AE Image Reconstruction}

\begin{figure}[!t]
\centerline{\includegraphics[width=0.9\columnwidth]{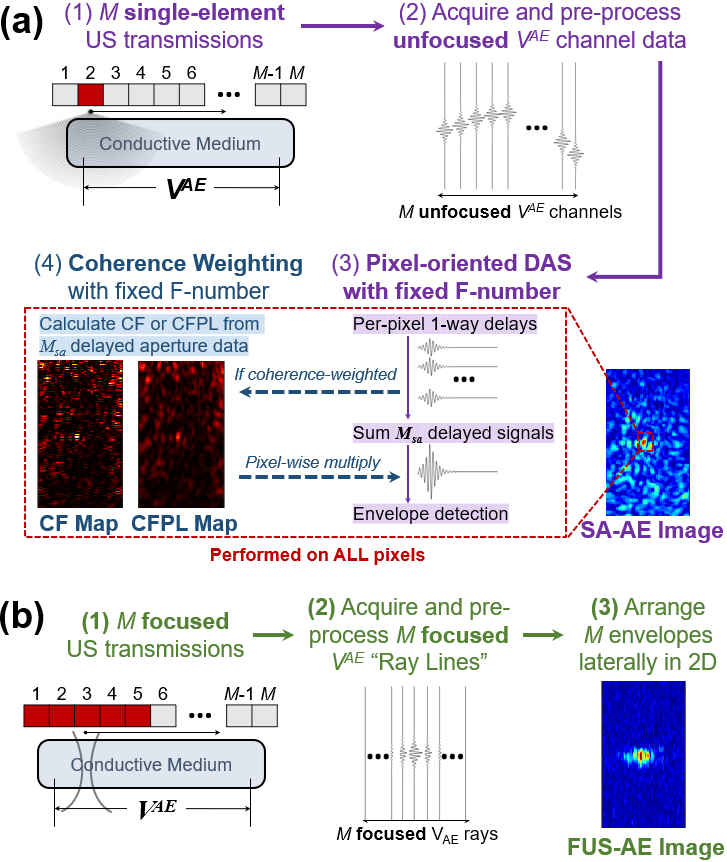}}
\caption{Outline of AE image reconstruction procedures. (a) Reconstruction steps for SA-AE, which mainly involved pixel-oriented DAS of unfocused $V^{AE}$ channel data with a fixed F-number. The CF and CFPL maps were calculated following \eqref{eq:CF} and \eqref{eq:CFPL} in cases where coherence weighting was applied. (b) Reconstruction steps for FUS-AE, which was a direct arrangement of the acquired AE ray lines into a 2D image. The signal envelopes were computed for image display in both methods.} 
\label{fig_recon}
\end{figure}

An AE image pixel grid was first defined for reconstruction. A sampling interval of 0.43$\lambda$ (same as the array element pitch) and 0.25$\lambda$ were used for the $x$ and $z$ directions, respectively. The field-of-view spanned the transducer width of 20.16 mm laterally, and 50 mm in depth to cover the full extent of the conductive zones. Fig. \ref{fig_recon}(a) shows the image reconstruction procedure for SA-AE.  An SA-AE image was reconstructed through pixel-oriented DAS of the single-element unfocused AE signals. Pixel delays were computed based on the time needed to propagate across the geometric distance between the emitting element and the pixel location, at the medium SoS. A fixed F-number of $F$ = 1.5 was selected, summing only signals from the $M_{sa}$ elements covered by the depth-dependent sub-aperture, as described in \eqref{eq:M_sub}. The Hilbert transform envelope of the beamformed SA-AE data was then computed to obtain an SA-AE image. For coherence-weighted SA-AE, the beamformed data was pixel-wise multiplied with the CF \eqref{eq:CF} and CFPL \eqref{eq:CFPL} maps prior to envelope detection.

The reconstruction steps for FUS-AE are shown in Fig. \ref{fig_recon}(b). FUS-AE images were reconstructed by directly arranging the FUS-induced AE channel data laterally into a 2D image, and the temporal axis was mapped to the depth ($z$) axis according to the propagation time at the medium SoS. For comparable image evaluation, the FUS-AE beamformed data were downsampled to match the sampling interval defined for the SA-AE images. The Hilbert transform envelope was then computed to display an FUS-AE image.

\subsection{Experimental Setup}

\begin{figure}[!t]
\centerline{\includegraphics[width=\columnwidth]{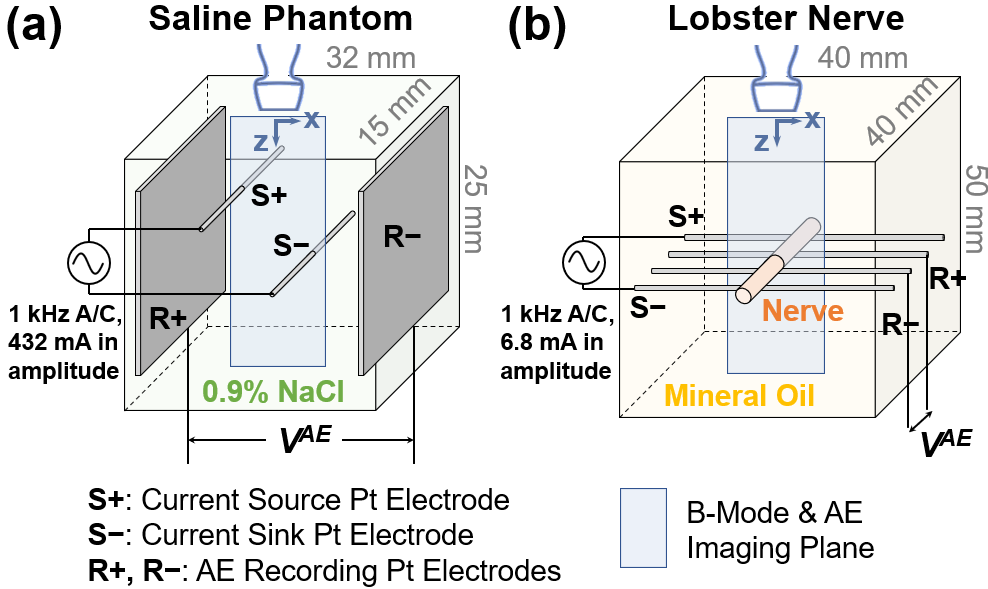}}
\caption{Schematics of conductive domains imaged. (a) The saline phantom consists of an insulated volume of 0.9\% NaCl solution, suspended in a tank of deionized water (not shown). A rubber sheath provided an acoustic window for insonification of the saline volume. (b) An \textit{ex vivo} lobster nerve cord was placed across four electrodes, separated at 5 mm, and bathed in mineral oil. The ends of the nerve cord (not shown) were soaked in Ringer's solution for tissue maintenance. In (a) and (b), the blue rectangle shows the plane for B-Mode and AE imaging. Note that the imaging field extends beyond the conductive regions shown, depending on the multiple depths at which the transducer was held.}
\label{fig_phantoms}
\end{figure}

Two conductive domains were imaged in this work. The first phantom (Fig. \ref{fig_phantoms}(a)) was a saline volume ($32~\text{mm}~(x)\times15~\text{mm}~(y)\times25~\text{mm}~(z)$) consisting of 0.9\% NaCl solution suspended in a tank of deionized water. A rubber sheath was used to provide an acoustic window into the NaCl solution, while maintaining electrical insulation between the conductive region and the surrounding deionized water. The receive electrodes used in the saline phantom were two Pt plate electrodes ($15~\text{mm}\times15~\text{mm}$) placed out of the acoustic field-of-view. For the saline phantom, an AC amplitude of 432 mA was supplied via two Pt wire supply electrodes, and AE signals were averaged across $k=2048$ transmits on the oscilloscope. The medium SoS was taken as 1480 m$\cdot$s$^{-1}$.
 
 The second medium (Fig. \ref{fig_phantoms}(b)) consisted of an \textit{ex vivo} lobster ventral nerve cord (7.5 mm length, 2 mm diameter), placed across two supply and two receive Pt wire electrodes spaced at 5 mm, with its ends submerged in Ringer's solution for maintenance. The nerve was surrounded by a mineral oil to electrically insulate it from the US probe. The imaging plane cut across the central transverse section of the nerve between the receive electrodes. An AC amplitude of 6.8 mA was supplied through the nerve, and AE signal measurements were averaged $k=128$ times. Rough measurements of the SoS in the mineral oil bath revealed a similar SoS to that in water, hence the medium SoS was also taken as 1480 m$\cdot$s$^{-1}$.

 Both media were imaged with the same US transducer array at several depths to simulate realistic imaging scenarios where the source locations were unknown prior to imaging. At all configurations, the focal depth in the FUS-AE scheme was fixed at $f_{FUS}=22$ mm.

\subsection{AE Image Evaluation Metrics}

For each medium, electrical sources were identified as the target for image evaluation. In the saline phantom, both supply electrodes served as targets; for the nerve cord, the signal source was the transverse nerve section. The spatial resolution was quantified by measuring the full-width at half-maximum of the axial and lateral profiles crossing the peak signal pixel for each source target in the enveloped AE images, to obtain the axial resolution (AR) and lateral resolution (LR), respectively. To measure the focusing quality achieved by each imaging strategy, the peak sidelobe level (PSL) was evaluated from the lateral profiles of the enveloped AE image as follows:
\begin{IEEEeqnarray}{c}
\text{PSL}=20\cdot\log_{10}(A_{S_{max}}/A_M),
\label{eq:PSL}
\end{IEEEeqnarray}
where $A_{S_{max}}$ and $A_M$ refer to the peak sidelobe amplitude and the main lobe amplitude, respectively. To investigate SNR performance, the signal-to-noise ratio (SNR) was computed from the pre-enveloped AE images as follows:
\begin{IEEEeqnarray}{c}
\text{SNR}=10\cdot\log_{10}\left(\frac{
1/N_S \sum_{i=1}^{N_S} x_i{}^2}{1/N_N \sum_{i=1}^{N_N} (y_i-\bar{y}){}^2}\right),
\label{eq:SNR}
\end{IEEEeqnarray}
where $x_i$ is the $i$-th pixel value within the target ROI (of $N_S$ pixels) and $y_i$ is the $i$-th pixel value within the background ROI (of $N_N$ pixels) without any electrical sources.

\section{Results}
\label{sec:results}

\subsection{SA-AE Provides More Uniform Reconstruction Quality than FUS-AE across Entire DOF}

\begin{figure}[!t]
\centerline{\includegraphics[width=0.75\columnwidth]{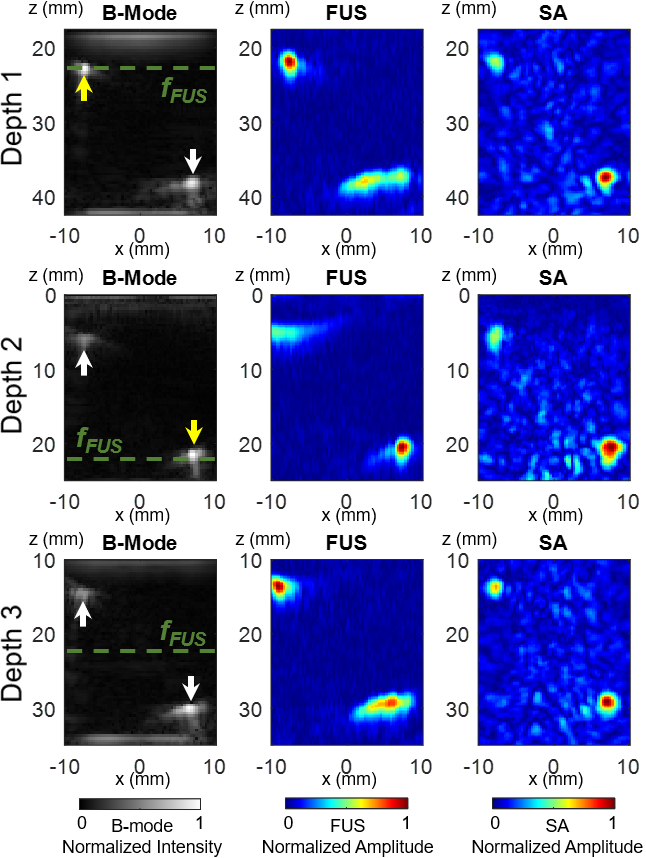}}
\caption{B-Mode and AE images acquired for the saline volume. FUS-AE and SA-AE imaging were performed with the phantom positioned at three different depths, while keeping the focal depth fixed at $f_{FUS}=22$ mm for FUS-AE. The S+ and S- electrodes (marked with yellow arrows) were located within the FUS-AE focal zones in the first and second depths, respectively. The remaining sources were not directly focused by FUS-AE (marked with white arrows). When properly focused (yellow arrow targets), FUS-AE imaging was able to reconstruct the point-like electrode sources satisfactorily. However, when out of focus, the sources were smeared. The electrodes retain their point-like geometry across all six positions in the SA-AE images. However, a background noise which increases with imaging depth was observed in the SA-AE images.}
\label{fig_salineMain}
\end{figure}

\begin{figure}[!t]
\centerline{\includegraphics[width=0.75\columnwidth]{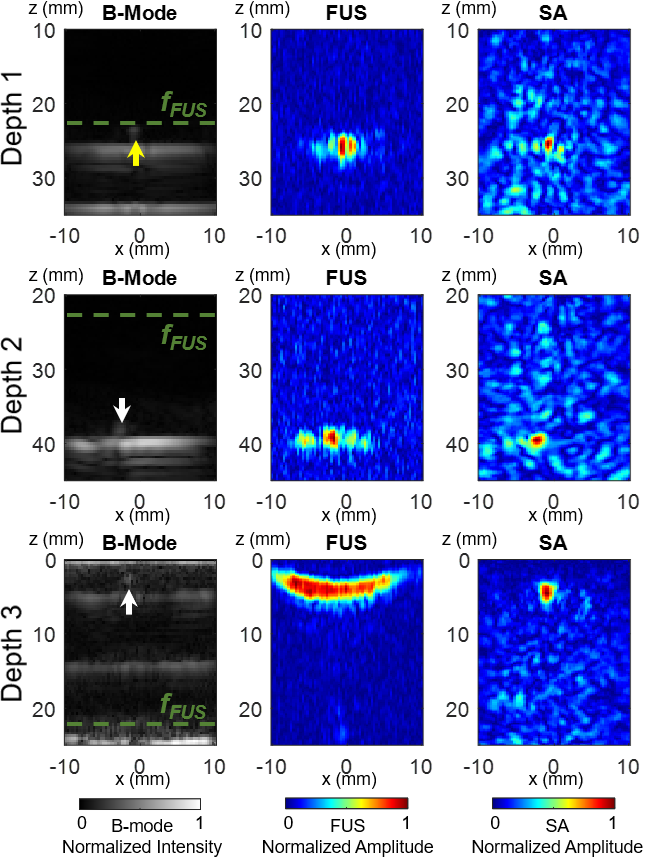}}
\caption{B-Mode and AE images acquired for the \textit{ex vivo} lobster nerve. FUS-AE and SA-AE images of the nerve were acquired at three different depths, while fixing the focal depth at $f_{FUS}=22$ mm for FUS-AE. The FUS waves were directly focused onto the nerve only in Depth 1 (marked with a yellow arrow), but not in Depths 2 and 3 (marked with white arrows). When the nerve was within the FUS focus (yellow arrow target), FUS-AE was able to reconstruct the circular geometry of the current-conducting nerve section, albeit with some side lobes. When out of focus (white arrow targets), the nerve section was smeared in the FUS-AE images. In the SA-AE images, the nerve sections retain their circular geometry. However, like in the saline phantom, a background noise which increases with imaging depth was observed.}
\label{fig_nerveMain}
\end{figure}

Fig. \ref{fig_salineMain} shows the B-Mode, FUS-AE and SA-AE images acquired for the saline phantom. Among the six positions at three depths, two of the electrode sources were within the FUS focus (marked with yellow arrows in Fig. \ref{fig_salineMain}). At these positions, the electrodes appeared as point-like sources in the FUS-AE images; whereas the remaining sources not under direct focus were smeared and weaker in amplitude. In the SA-AE images, all six electrode positions retained their point-like appearance without noticeable smearing. In the same image, the source in the deeper zone appeared stronger than the other source in the shallow region. Moreover, for the SA-AE images, a background noise component which increased with imaging depth was observed. The weaker sources were comparable in amplitude to the background noise, showing that the background noise degraded image contrast in the basic SA-AE imaging strategy. Similar observations were made for the lobster nerve section (Fig. \ref{fig_nerveMain}). FUS-AE could reconstruct the circular section of the nerve cord only when it was located at the chosen focal depth. Otherwise, the nerve was reconstructed with strong smearing. SA-AE imaging was able to reconstruct the circular geometry of the nerve transverse section at all depths. Like in the saline case, obvious background noise was present in the SA-AE images, leading to poor image contrast.

\subsection{Coherence-based SA-AE Reduces Background Noise}

\begin{figure}[!t]
\centerline{\includegraphics[width=\columnwidth]{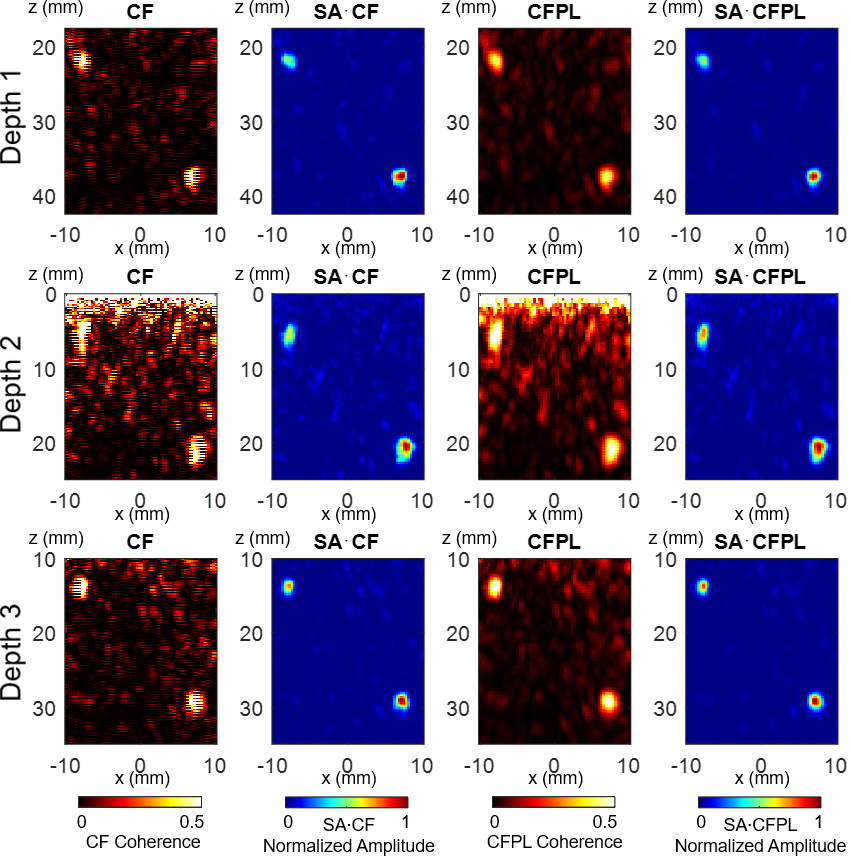}}
\caption{CF and CFPL weighted SA-AE images for the saline phantom. Depths 1-3 correspond to those in Fig. \ref{fig_salineMain}. The CF and CFPL maps exhibit higher coherence near the electrode sources. Upon application of the CF and CFPL weightings, the weighted SA-AE images showed clear point-like electrode sources without noticeable background noise, unlike in the basic SA-AE images in Fig. \ref{fig_salineMain}. The CFPL maps were smoother and more continuous than the CF maps, resulting in smoother CFPL-weighted SA-AE images.}
\label{fig_salineCF}
\end{figure}

\begin{figure}[!t]
\centerline{\includegraphics[width=\columnwidth]{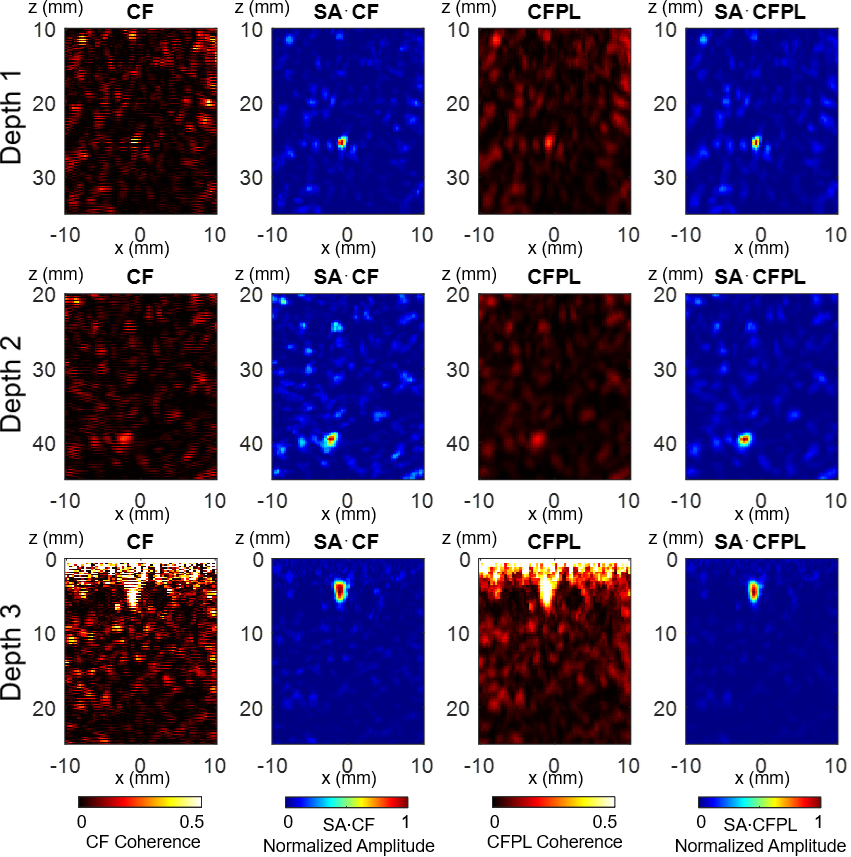}}
\caption{CF and CFPL weighted SA-AE images for the \textit{ex vivo} lobster nerve. Depths 1-3 correspond to those in Fig. \ref{fig_nerveMain}. The CF and CFPL maps exhibit higher coherence near the nerve section. Applying the CF and CFPL weightings resulted in cleaner SA-AE images with reduced background noise, clearly revealing the nerve section region. The CFPL maps were smoother and more continuous than the CF maps, resulting in smoother CFPL-weighted SA-AE images. In particular, for Depth 2, the background noise suppression was stronger for CFPL than CF.}
\label{fig_nerveCF}
\end{figure}

For the saline phantom, the CF and CFPL maps computed revealed higher coherence near the electrodes (Fig. \ref{fig_salineCF}). Upon application of the coherence weightings, the background noise in the weighted SA-AE images was suppressed compared to the basic reconstruction in Fig. \ref{fig_salineMain}. Consequently, clearer point electrode sources were reconstructed in the weighted SA-AE images. Compared to CF, the CFPL map showed a smoother and more continuous coherence distribution, resulting in a smoother weighted SA-AE image. Fig. \ref{fig_nerveCF} shows similar results for the nerve, where the AE signal components showed higher CF and CFPL values at the nerve transverse section. The weighted SA-AE images revealed a cleaner, localized nerve section as the only electrical source in the imaging field. The background noise was strongly suppressed compared to the basic SA-AE images in Fig. \ref{fig_salineCF}. Like in saline, the CFPL maps were smoother and more continuous, resulting in cleaner noise suppression than the CF maps. This is particularly apparent for Depth 2 in Fig. \ref{fig_nerveCF}, where the background noise was lower for the CFPL-weighted image.

\subsection{Quantitative Results from Evaluation Metrics}

\begin{figure}[!t]
\centerline{\includegraphics[width=\columnwidth]{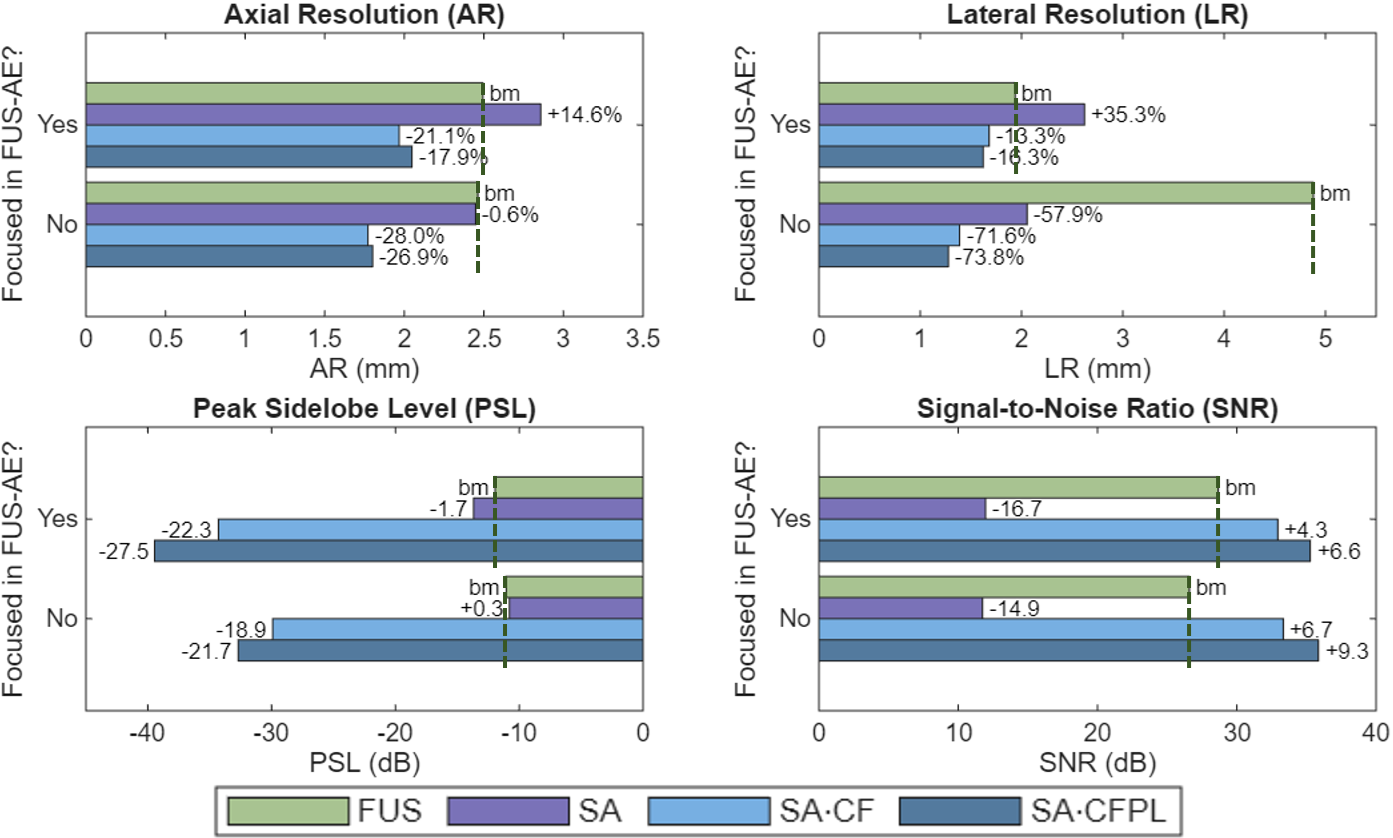}}
\caption{The evaluation was performed separately for electrode sources which were on-focus vs. off-focus in the FUS-AE scheme. The mean value for each metric was reported. The percentages (\%) and numbers (in dB) adjacent to each bar showed the change in the corresponding metric relative to FUS, which served as the benchmark (marked with ``bm") within each sub-group. The spatial resolution, AR and LR, generally improved for the SA-based methods compared to FUS, particularly for the off-focus targets. In most cases, PSL was lower for the SA-based methods. SNR decreased sharply for SA compared to FUS; however the CF and CFPL-weighted SA boosted SNR, showing a large increase in SNR from FUS.}
\label{fig_salineMetrics}
\end{figure}

\begin{figure}[!t]
\centerline{\includegraphics[width=\columnwidth]{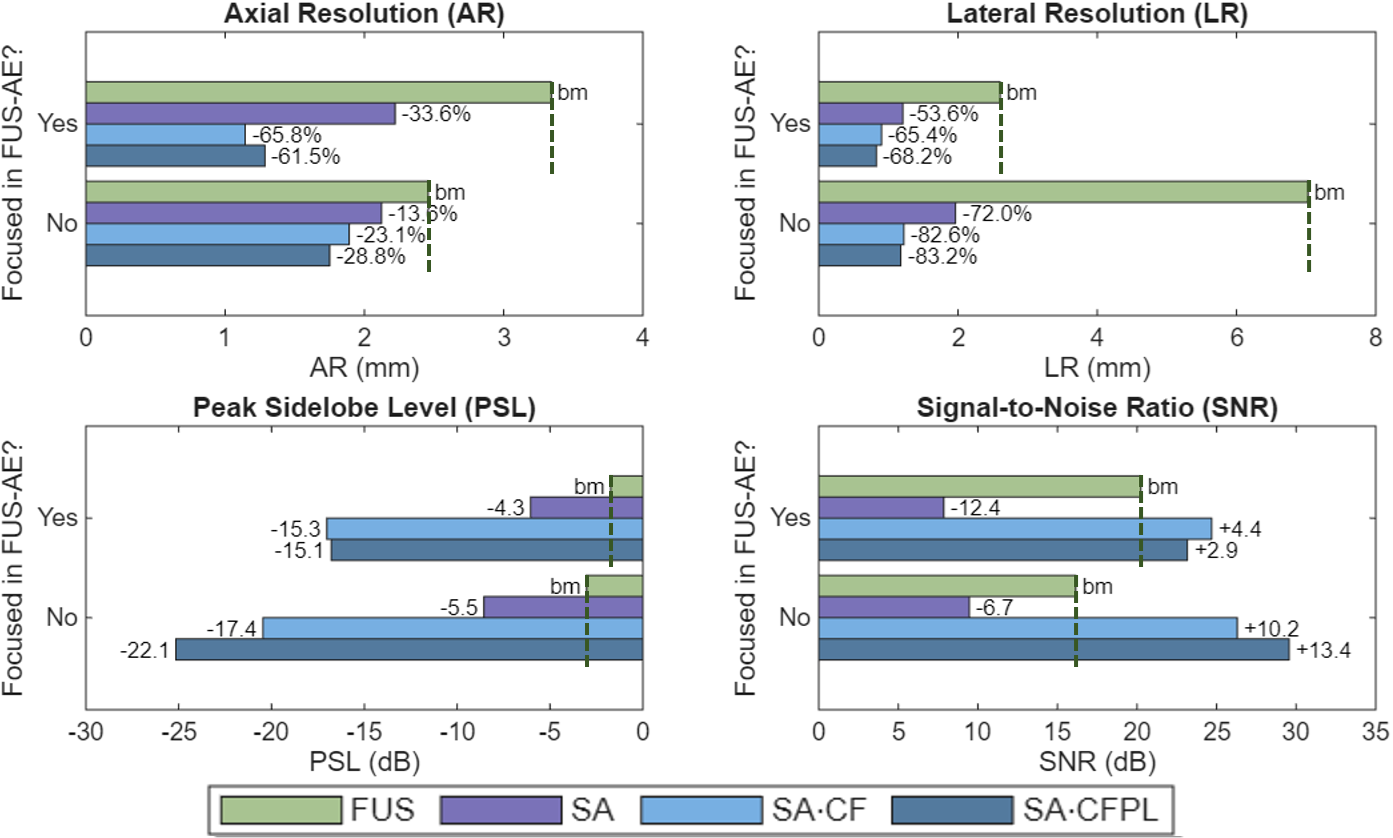}}
\caption{The evaluation was performed separately for electrode sources which were on-focus vs. off-focus in the FUS-AE scheme. The mean value for each metric was reported. The percentages (\%) and numbers (in dB) adjacent to each bar showed the change in the corresponding metric relative to FUS, which served as the benchmark (marked with ``bm") within each sub-group. Spatial resolution (AR and LR) and PSL was improved for the SA-based methods, especially for the off-focus cases. SNR dropped strongly for SA; however CF and CFPL boosted SNR considerably, showing a large increase in SNR from FUS.}
\label{fig_nerveMetrics}
\end{figure}

Figs. \ref{fig_salineMetrics} and \ref{fig_nerveMetrics} the evaluation metrics computed from the AE images of the saline phantom and lobster nerve, respectively. Each medium contained electrical sources that may not have been directly focused by the FUS-AE imaging scheme, depending on the depth at which the transducer array was held. The metrics were compared across the imaging techniques, separately in cases of on-focus and off-focus electric sources.

Both experiments reported similar trends in the computed metrics. In terms of spatial resolution, SA-AE showed some deterioration (for saline case) and marginal improvement (for nerve case) in AR and LR compared to FUS-AE when the target was at the FUS focal depth. However, for the off-focus targets outside the FUS focal zone, AR and LR degraded for FUS-AE; whereas spatial resolution was relatively stable for SA-AE. In most cases, SA-AE also exhibited a lower PSL than FUS-AE, indicating a lower sidelobe level and hence higher focusing quality for SA-AE. However, the uniform spatial resolution and improved focusing of SA-AE was accompanied by a decrease in SNR, for both on-focus and off-focus sources. The further application of CF and CFPL weightings onto the SA-AE images resulted in an overall improvement across all metrics. CF and CFPL saw a further reduction in AR, LR and PSL compared to the basic SA method. Moreover, the SNR for CF-weighted and CFPL-weighted SA-AE images increased compared to FUS-AE. Among the two weightings, the extent of metric improvement was stronger for CFPL than CF.

\section{Discussion}
\label{sec:discussion}

In this study, we first approached the AE imaging reconstruction problem from a beamforming perspective. We emphasized that the transmit beam pattern $b$ serves as the spatial siftor of the $s$ field in AE imaging. The effects on AE imaging performance brought upon by various transmission schemes can be explained by looking at the effective beam patterns used. Through the design and recombination of transmitted waves, the beam pattern can be tuned to maximize AE image quality in various ways to meet application-specific demands. Suppose that the target application involves sources distributed across the DOF (as in the saline phantom), or if their actual location is hard to determine from B-Mode imaging due to low acoustic contrast (as in the lobster nerve). Using such scenarios as a case study, we demonstrated the effects of using different beam patterns on the AE $s$ field reconstruction.

\subsection{When SA-AE Adds Value}

When unknown source depths are involved, FUS waves may result in poor spatial resolution at sources located outside its focus. The white arrow targets in Figs. and \ref{fig_salineMain} and \ref{fig_nerveMain} highlight these scenarios. In the FUS-AE images, the targets smeared according to the diverging or converging wavefronts of the transmitted FUS waves, when they were in the far field or near field of the focused beam, respectively. In such cases, the FUS-AE images were unable to localize electrical sources accurately. A multi-foci approach may be useful; however, it may be difficult to decide on the number of foci needed if the number of sources and their depth locations are unknown.

In such scenarios where electrical sources are positioned arbitrarily throughout the DOF, a beam pattern which provides a uniformly pixel-specific sifting ability would be more appropriate. We showed in \eqref{eq:shat_SA}-\eqref{eq:shat_SA_approx} that an SA approach, which performs DAS using AE signals induced by unfocused waves, is able to dynamically synthesize a pixel-specific focused beam siftor. Using single-element transmissions, the SA-AE strategy was implemented and compared alongside FUS-AE images on two conductive domains. The direct implementation of SA-AE results in reconstructed targets which were relatively localized, regardless of their depth positions. The reconstructed sources, i.e. the electrodes in the saline phantom and the circular nerve cross section, showed closer alignment with their true positions shown in the B-Mode images, compared to those in FUS-AE.

One key advantage of SA-AE is the post-acquisition flexibility in using the signal channels, since they were independently acquired with decoupled acoustic sources. In our work, we leveraged this aspect to implement a dynamic sub-aperture in the reconstruction of SA-AE images to maintain uniform focusing power throughout the DOF. Beyond this, other reconstruction strategies such as varying the effective transmit apodization \cite{allard2025ultrasound} or performing phase aberration corrections \cite{preston2023acoustoelectric} could be realized using such SA-AE channel data.

\subsection{Further Beamforming Supplements Practical SA-AE}

Despite the resolution improvement and reconstruction flexibility brought by SA-AE, it is accompanied by a drop in SNR due to the inherently weak AE signal nature exacerbated by the low acoustic pressure of the unfocused spherical waves. Moreover, due to additive noise, DAS in SA-AE intrinsically causes a drop in SNR compared to FUS-AE at the same level of averaging, as described by \eqref{eq:SNR_SAFUS}. Although FUS-AE enjoys a higher SNR level, it uses higher acoustic pressures associated with beam focusing. Meanwhile, SA-AE synthesizes dynamic foci, without applying the same focused pressure levels. In our work, a comparison between FUS-AE and SA-AE using the same maximum spatial pressure level was not performed. Such a study may elucidate whether: (1) SA may present as a safer alternative to FUS for establishing large effective pressure levels needed for AE imaging, and (2) SA still suffers a lower SNR than FUS under the same applied pressure. Additionally, the use of large transmit apertures in FUS-AE results in large electromagnetic inference (EMI) signals which corrupt the immediate AE temporal signal samples during the onset of ultrasound transmission. This may distort the reconstruction of electric sources in the ultrasound near field, e.g., the nerve section in Depth 3 (Fig. \ref{fig_nerveMain}). The smaller EMI signals from single-element transmissions in SA-AE may result in a smaller disruption to the near field AE image quality.

Moreover, the use of unfocused waves in SA-AE results in a diffuse modulation of the conductive medium. Equation \eqref{eq:shat_SA} indicates that the DAS step at each point includes a certain degree of signal contribution from the other iso-temporal points. This means that electrical sources may ``leak" signal into other points, possibly resulting in the structured appearance of the reconstructed background noise. To investigate this hypothesis, we additionally performed a sham experiment on the saline phantom, where the US array was blocked during AE acquisition. Reconstructed SA-AE images revealed a similar structured background noise pattern (Supplementary Fig. \ref{fig_shamnoise}), indicating that the electrical noise from our instrumentation was the main source of these patterns, rather than the leakage of AE modulation signals. Related to this, while our current SA-AE results show a reduced SNR performance compared to FUS-AE, we envision future hardware improvements would lower the noise floor in AE signal measurements, therefore reducing the averaging demand and improving the feasibility of SA-AE when it comes to imaging weak electric sources.

Nonetheless, under the SNR limits of our existing instrumentation, we demonstrated that the imaging performance of SA-AE can be improved by supplementing the technique with spatial coherence-based weighting. The coherence-based weightings using the CF and CFPL maps approached the SA-AE reconstruction problem from two aspects. Firstly, the possibility of signal leakage from DAS that results in image artifacts, such as side lobes, is still of concern. Like conventional CF for B-mode imaging \cite{hollman1999coherence}, coherence is highest only when the AE signals induced for a given source are aligned at its true position. For the surrounding pixels, any leaked signal is accompanied by fewer coherent channels in the delayed aperture and can therefore be suppressed by the lack of spatial coherence across the aperture. This explains why the CF and CFPL-weighted SA-AE images exhibited further reductions in LR and PSL, because of enhanced suppression of side-lobes from coherence filtering. Secondly, the SA-AE background noise component was found to decrease with averaging (Supplementary Fig. \ref{fig_shamnoise}). This feature indicates that the noise was incoherent and therefore should not exhibit spatial coherence across the delayed aperture. Instead of relying on extensive averaging, coherence maps help accelerate noise suppression using a lower number of total transmissions by capturing the spatial incoherence of electronic noise signals across the aperture. While CF has usually been adopted for enhancing contrast through sidelobe reduction for relatively high SNR conditions, it has been shown to widen the gap between the signal and noise floor under low SNR signal regimes \cite{nilsen2010wiener}. Applying CF on low-SNR signals effectively boosts their SNR, albeit with some suppression on the desired signals themselves. Further extensions to this technique may incorporate the Wiener postfilter \cite{nilsen2010wiener} to balance between the robustness and filtering performance of coherence weighting.

The electric sources in the CF and CFPL-weighted images were visibly finer than those in the FUS-AE and basic SA-AE images, in particular for the nerve cord. Considering the 2 mm diameter of the nerve section, the LRs of roughly 1 mm obtained in the coherence-weighted images appear to present an underestimate of the size extent of the conductive zone. As there was a tissue-oil interface at the boundary of the nerve tissue, phase aberrations may have sharply degraded spatial coherence, resulting in suppression of the AE signal near the boundaries of the nerve section and therefore the appearance of a narrower nerve section. Between CF and CFPL, CFPL exhibited a stronger suppressive effect as it fully utilized the data samples spanning the temporal extent of the transmitted pulse lengths. As such, CFPL showed stronger filtering power in its coherence values, resulting in a greater reduction in LR, and the added benefit of stronger noise suppression than CF.

\subsection{Further Comments on SA-AE Reconstruction}

At this point, we remind the reader that the crux of AE imaging is a measurement of the $\hat{s}$ field (see \eqref{eq:shat_x1_approx}), to obtain a mapping of the true $s$ distribution (see \eqref{eq:s}). While mapping $\hat{s}$ from AE channel data is practically instant for FUS-AE, SA-AE image reconstruction requires additional computational time for DAS. In our basic CPU-based implementation, the reconstruction time for one SA-AE image frame is roughly 22 s, with an additional 12 s needed for computing the CF and CFPL maps, giving a reconstruction frame rate of 0.03 Hz. GPU-based sparse matrix beamforming has been shown to boost RF DAS beamforming by at least 100-fold \cite{hou2014sparse}. Such strategies are directly applicable to SA-AE, to allow fast visualization of SA-AE image frames. For an imaging depth of 50 mm at a SoS of 1480 m$\cdot$s$^{-1}$, the ideal acquisition rate (without repeated transmissions) for a complete set of 64-channel AE data is 462.5 Hz. This figure quickly drops with an increase in repeated transmissions for signal averaging. Considering that the basic SA-AE technique is prone to low SNR and therefore requires extensive averaging, the true bottleneck in real-time SA-AE based imaging would be the averaging demand, rather than the reconstruction time incurred.

\begin{figure}[!t]
\centerline{\includegraphics[width=0.8\columnwidth]{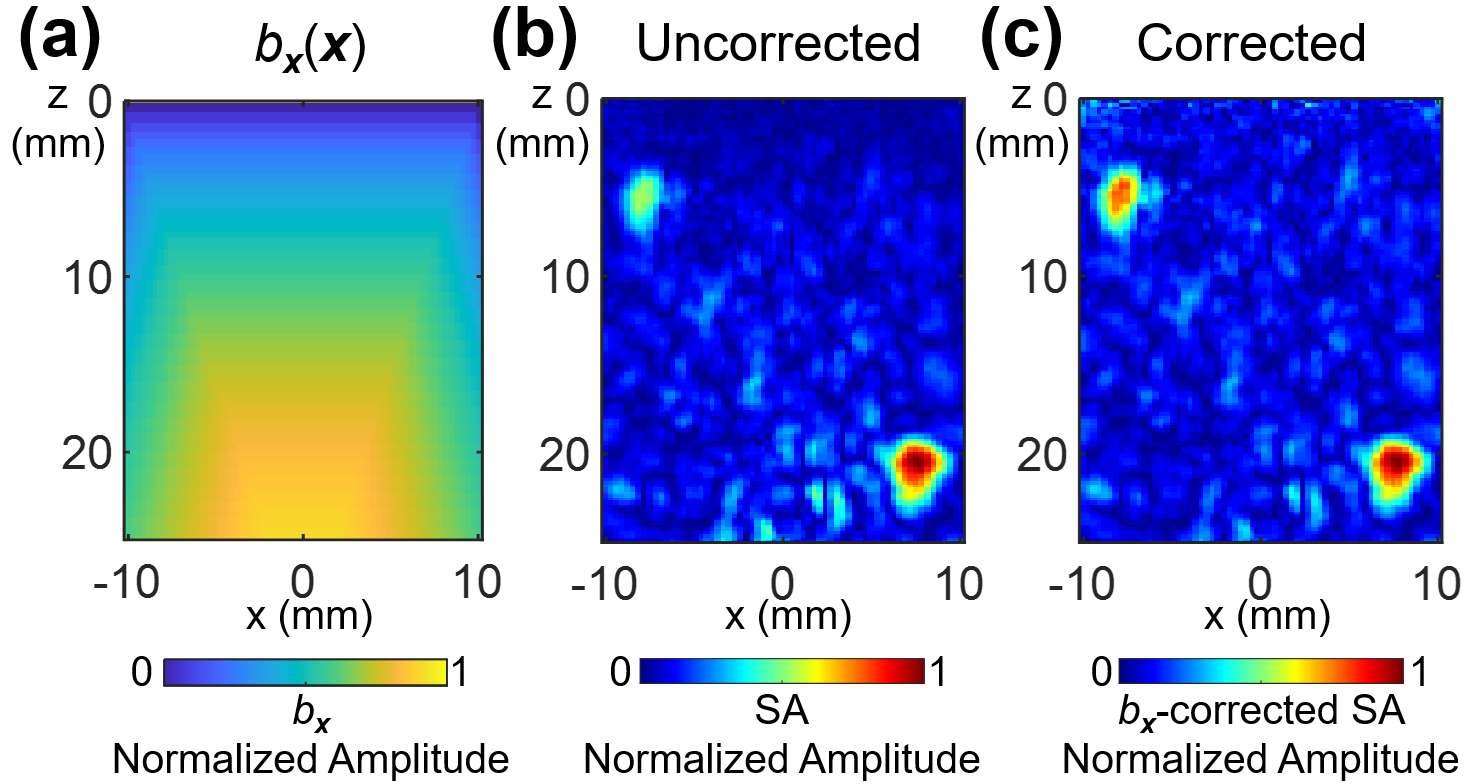}}
\caption{Image correction with effective beam weighting function. (a) Effective beam weighting function $b_\textbf{x}(\textbf{x})$ used for SA-AE reconstruction. The beam amplitude was increased with depth due to the increasing size of transmit sub-aperture used, and decreased with increasing lateral displacement due to the finite size of the transducer footprint. (b) Uncorrected SA-AE image, showing a larger deviation in amplitude between the two electrode sources. (c) $b_\textbf{x}(\textbf{x})$ corrected SA-AE image, depicting a more similar amplitude between the two electrodes passing the same level of current.}
\label{fig_bcorrection}
\end{figure}

A direction to address averaging demand is by applying coded excitation schemes. Chirp excitation \cite{qin2012mapping} and temporal encoding using cascaded dual-polarity waves (CDW) \cite{tang2025signal} have been shown to enhance AE SNR. As a supplementary study (Supplementary Fig. \ref{fig_nerveCDW}), we employed CDW waves to conduct SA-AE imaging, while benchmarking it against the single-pulse SA-AE presented in this work. We demonstrated that when SA-AE is used in tandem with temporal encoding, its SNR could be further boosted to approach that of FUS-AE (Supplementary Table \ref{tab_nerveCDW}). This would allow SA-AE to perform $s$ field reconstructions at an increased frame rate, pushing it toward capturing transient changes in biological electric fields, such as action potential propagation in nerve tissue. CDW encoding is uniquely suitable for AE imaging as the issue of a large dead zone \cite{zhang2017ultrafast} associated with extended CDW transmissions is eliminated, since AE signals are measured by electrodes and not by the transducer array. Longer CDW pulses could be used to further boost AE SNR, as long as the EMI signal is not too large to corrupt AE signal content. Single-element transmissions in SA-AE result in weaker EMI, making SA-AE a good fit with extended CDW pulses.

In SA-AE images, an increase in the $\hat{s}$ field amplitude was observed with increasing depth. For example, the S+ electrode located in the shallower region (Figs. \ref{fig_salineMain} ad \ref{fig_salineCF}) showed a weaker signal than the S- electrode, although both were expected to show similar intensities as they passed the same level of current. This amplitude distribution resulted from the dynamic sub-aperture adopted, which summed fewer signal channels in the shallow zone during DAS. In fact, while the SA-synthesized beam weighting functions $b_{\textbf{x}}$ were spatially specific at each pixel, they did not exhibit similar pressure intensity across the DOF. To depict the $s$ field more faithfully, $\hat{s}$ should be amplitude-corrected by $b_{\textbf{x}}$ (Fig. \ref{fig_bcorrection}). In general AE imaging, we similarly recommend correction of the acquired $\hat{s}$ maps by the effective beam weighting function, especially when the relative intensity of sources is important or when the applied beam spatial intensities are highly non-uniform.

\subsection{Study Limitations and Future Work}

We demonstrated the SA-AE framework using conductive media that have homogeneous background SoS. In the \textit{in vivo} setting, inhomogeneous tissue layers such as adipose tissue present phase aberrations that degrade the focusing quality and hence imaging performance for both FUS-AE and SA-AE. With cross-correlation-based identification of a source reference's signals across the AE aperture data, time-reversal of AE signals has been demonstrated to refine the synthesized focus in the presence of phase aberrators \cite{preston2023acoustoelectric}. Moreover, like in B-Mode imaging \cite{li2003adaptive}, the coherence weightings shown in our work would likely help suppress the focusing errors caused by aberration layers. Our future work is to further validate the use of SA-AE imaging in \textit{in vivo} nerve tissue, to determine its feasibility in imaging transient biological current flow, surrounded by inhomogeneous tissue media.

As DAS reconstruction of SA-AE is computed based on the medium SoS, inaccuracies in the presumed SoS would result in distorted image reconstructions \cite{perrot2021so}. Since AE imaging can be performed simultaneously with B-mode ultrasound, SoS estimation and correction strategies could be adapted to reconstruct a medium SoS map. For example, B-mode segmentation-guided SoS-aware beamforming \cite{de2025layer} has been shown to improve DAS-based reconstruction in layered inhomogeneous SoS media. The estimated SoS maps could then be used for SoS-aware SA-AE imaging. Since AE channel data consists of RF hyperbolic signatures like in ultrasound, the AE data itself could also be used directly to infer an average SoS for image reconstruction. For instance, the suitable SoS estimate could be determined by finding one that minimizes the phase dispersion along signal hyperbolas \cite{perrot2021so}. Depending on the severity of the SoS inhomogeneity, SoS correction may be an integral aspect in the realization \textit{in vivo} SA-AE imaging.

While we performed measurements and reconstructions on one lead pair measurement, the techniques discussed are readily extensible to multiple-lead setups for electric vector field reconstructions. 2D and 3D electric vector field mappings would require at least two and three orthogonal leads, respectively \cite{olafsson2008ultrasound}. In this work, although the transmitted wavefields were actually propagating in 3D, the 1D P4-2 array mainly insonified the central $xz$-plane due to its built-in acoustic lens. Should 3D mapping of the electric fields be of interest, 2D matrix arrays \cite{allard2024neuronavigation} would provide greater flexibility in synthesizing the AE sifting beams in 3D space.

\section{Conclusion}

In this work, we first dissected the AE imaging problem by looking at the role of the US beam pattern in the image formation process. The beam acts as the weighting function which determines the modulation region for the AE effect, therefore enabling localized probing of electric fields by US waves. We show theoretically and experimentally that such a sifting beam can synthesized from unfocused waves through a SA strategy, in contrast from conventional approaches which use FUS waves. The SA approach provides flexible combination of decoupled AE signal channels into the final AE image, enabling pixel-based focusing and dynamic tuning of the transmit F-number. Compared to FUS-AE, SA-AE generally improved spatial resolution across the DOF, but suffered from strong background noise, which was an intrinsic limitation of SA-AE due to thermal noise amplification by its DAS operation. Coherence-based weighting with SA-AE was shown to provide some SNR recovery, closing the SNR gap between FUS and SA; while providing further resolution and contrast improvement via sidelobe reduction. The SA-AE beam weighting function was further used to correct the amplitude of the acquired AE image, to raise the importance of such considerations toward the accurate mapping of the electric fields in AE imaging. The overall goal of this work was to emphasize the importance of viewing the AE imaging problem from this beam-shaping perspective, thereby providing insights into the design process of suitable transmission schemes for application-based AE imaging.

\section*{Acknowledgment}

The authors wish to thank Prof. Jean Provost for his expert advice regarding the instrumentation amplifier selection, Prof. Matthew O'Donnell and Prof. Russell Witte for their insightful discussions and comments on the noise analysis.

\bibliographystyle{IEEEtran}
\bibliography{ref}

\newpage
\section*{Supplementary Material}

\subsection{Reconstructed Noise in SA-AE Sham Experiment}
\renewcommand{\thefigure}{S1.\arabic{figure}}
\setcounter{figure}{0}
\renewcommand{\thetable}{S1.\arabic{table}}
\setcounter{table}{0}

The experiments in the saline phantom were conducted with $k=\{512,1024,2048\}$ averages to measure the SA-AE signal. Under the same transducer configuration, a sham experiment was performed by blocking the transducer array was blocked with acoustically absorbent material. The electrode sources were reconstructed in the regular saline experiment, but not in the sham experiment since acoustic modulation was absence. However, a similar background noise structure was reconstructed in the SA-AE images regardless of the presence of actual acoustic modulation. The source of this noise should be from the instrumentation, since the supply currents were continuously pumped and voltage measurements were taken in both experiments. The diminishing amplitude in background noise with increase in $k$ suggested that it was an incoherent noise component contained within the AE channel data.

\begin{figure}[!b]
\centerline{\includegraphics[width=\columnwidth]{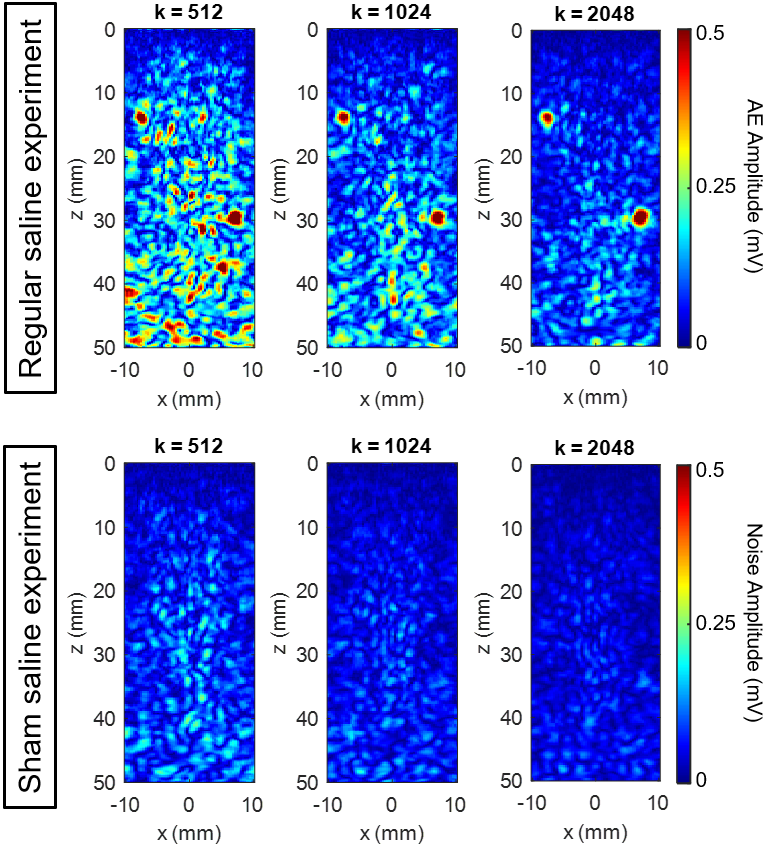}}
\caption{SA-AE reconstuction of a regular saline experiment (top row) and a sham experiment with the transducer blocked by acoustically absorbent material (bottom row). The background noise component was present regardless of actual medium insonification. The SA-AE images were acquired at increasing number of $k=\{512,1024,2048\}$ averages, showing that the noise was incoherent, as its magnitude decreased with more averaging.}
\label{fig_shamnoise}
\end{figure}

\newpage
\subsection{Temporal Encoding of SA-AE with CDW}
\renewcommand{\thefigure}{S2.\arabic{figure}}
\setcounter{figure}{0}
\renewcommand{\thetable}{S2.\arabic{table}}
\setcounter{table}{0}

\begin{figure}[!b]
\centerline{\includegraphics[width=\columnwidth]{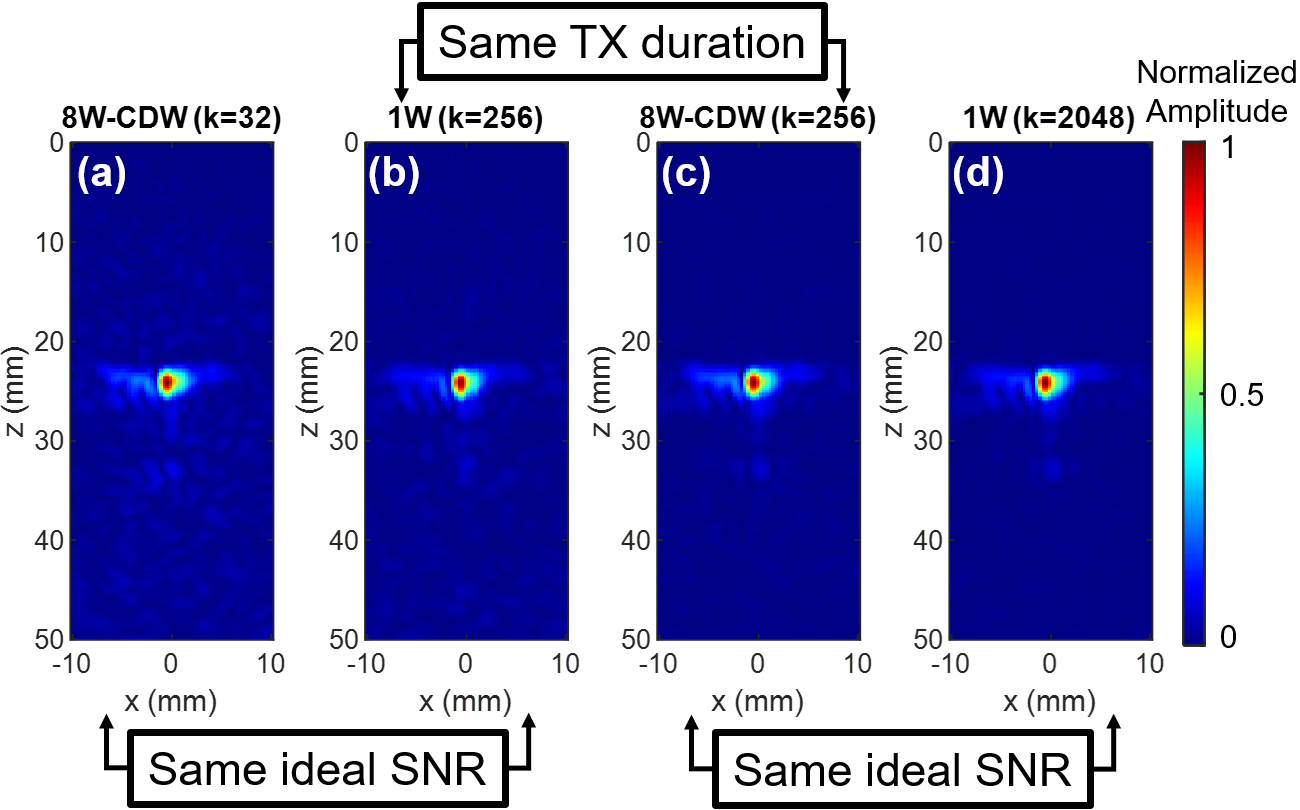}}
\caption{SNR enhancement of SA-AE images with CDW transmissions. SA-AE image acquired with: (a) 8-wave CDW with $k=32$; (b) single-wave with $k=256$; (c) 8-wave CDW with $k=256$; (b) single-wave with $k=2048$. $k$ refers to the number of repeated transmissions (TX). Note that the CDW-enhanced images have roughly $8\times$ the peak amplitude of the single-wave SA-AE images. The colorbars were clipped at a lower level to show the weaker background noise component.}
\label{fig_nerveCDW}
\end{figure}

\begin{table}[!b]
\caption{SNR for CDW and Reference SA-AE Images}
\label{table}
\setlength{\tabcolsep}{3pt}
\begin{tabular}{ccc|ccc}
\hline\hline
Method & k    & SNR (dB) & Method & k   & SNR (dB) \\
\hline
\multirow{2}{*}{1W}     & 256  & 30.6     & \multirow{2}{*}{8W-CDW}    & 32  (\textbf{-87.5\%}) & 29.7 (-0.9)    \\
                        & 2048 & 40.5     &                            & 256 (\textbf{-87.5\%}) & 37.8 (-2.7)    \\
\hline\hline
\multicolumn{6}{p{251pt}}{SNR comparison between SA-AE images (shown in Fig. \ref{fig_nerveCDW}) acquired using 8-wave CDW (8W-CDW) and single-wave (1W) reference. 8W-CDW achieved a similar (but slightly lower) SNR, but using only 12.5\% of the $k$ repetitions needed. Under the same $k=256$, 8W-CDW saw a 7.2 dB SNR increase (37.8 dB vs. 30.6 dB) compared to 1W SA-AE. The bolded percentages indicate a favourable drop in repetitions needed, corresponding to an increased imaging frame rate in practice.}
\end{tabular}
\label{tab_nerveCDW}
\end{table}

SA-AE images of the lobster nerve were further acquired with 8-wave CDW (8W-CDW) transmissions, with $k=32$ (Fig. \ref{fig_nerveCDW}(a)) and $k=256$ (Fig. \ref{fig_nerveCDW}(c)) repeated transmissions. These acquisitions were complemented with standard SA-AE images acquired using single-wave (1W) transmissions under the same $k$ repeats. From visual inspection of the background noise, 8W-CDW achieved similar SNR performances as their equivalent-SNR 1W counterparts, while using only 12.5\% of the $k$ transmissions needed. The computed SNR in Table \ref{tab_nerveCDW} revealed that 8W-CDW achieved a similar but slightly lower SNR. Under the same acquisition duration, i.e., same $k=256$, 8W-CDW achieved a SNR of 37.8 dB, showing a 7.2 dB increase in SNR compared to its 1W counterpart at 30.6 dB.

Overall, temporal encoding through CDW transmissions in SA-AE was effective in: (1) increasing AE image SNR under the same number of transmissions, (2) lowering the number of averages needed to achieve a similar SNR performance as the 1W version. This means that when SA-AE is implemented with CDW transmissions, its imaging frame rate can be enhanced significantly since less temporal averages are needed to measure signals with sufficient SNR.

\end{document}